\documentclass[journal]{IEEEtran}

\usepackage[utf8]{inputenc} 
\usepackage{url}            
\usepackage{booktabs}       
\usepackage{amsfonts}       
\usepackage{nicefrac}       
\usepackage{microtype}      
\usepackage{color}
\usepackage{lipsum}
\usepackage{graphicx}
\usepackage{siunitx}
\usepackage{amsmath}
\usepackage{multirow}
\usepackage{tikz}

\makeatletter
\let\MYcaption\@makecaption
\makeatother

\usepackage[font=footnotesize]{subcaption}

\makeatletter
\let\@makecaption\MYcaption
\makeatother

\begin{document}
\bstctlcite{IEEEexample:BSTcontrol}

\title{Results of the 2020 fastMRI Challenge for Machine Learning MR Image Reconstruction}
%
%
%
\author{Matthew~J.~Muckley$^{*,1}$,~\IEEEmembership{Member,~IEEE,}
        Bruno~Riemenschneider$^{*,2}$,
        Alireza~Radmanesh$^2$,
        Sunwoo~Kim$^3$,~\IEEEmembership{Member,~IEEE,}
        Geunu~Jeong$^3$,
        Jingyu~Ko$^3$,
        Yohan~Jun$^4$,
        Hyungseob~Shin$^4$,
        Dosik~Hwang$^4$,
        Mahmoud~Mostapha$^5$,
        Simon~Arberet$^5$,
        Dominik~Nickel$^6$,
        Zaccharie~Ramzi$^{7,8}$,~\IEEEmembership{Student Member,~IEEE,}
        Philippe~Ciuciu$^7$,~\IEEEmembership{Senior~Member,~IEEE,}
        Jean-Luc~Starck$^8$,
        Jonas~Teuwen$^9$,
        Dimitrios~Karkalousos$^{10}$,
        Chaoping~Zhang$^{10}$,
        Anuroop~Sriram$^{11}$,
        Zhengnan~Huang$^2$,
        Nafissa~Yakubova$^1$,
        Yvonne~W.~Lui$^2$,
        and~Florian~Knoll$^2$,~\IEEEmembership{Member,~IEEE}
\thanks{$^*$Equal contribution.}
\thanks{$^1$Facebook AI Research, New York, NY, USA}
\thanks{$^2$NYU School of Medicine, New York, NY, USA}
\thanks{$^3$AIRS Medical, Seoul, South Korea}
\thanks{$^4$Yonsei University, Seoul, Korea}
\thanks{$^5$Siemens Healthineers, Princeton, NJ, USA}
\thanks{$^6$Siemens Healthcare GmbH, Erlangen, Germany}
\thanks{$^7$CEA (NeuroSpin) \& Inria Saclay (Parietal), Université Paris-Saclay, F-91191 Gif-sur-Yvette, France}
\thanks{$^8$Département d’Astrophysique, CEA-Saclay, 91191 Gif-sur-Yvette, France}
\thanks{$^9$Radboud University Medical Center, Nijmegen, Netherlands}
\thanks{$^{10}$Amsterdam UMC, Amsterdam, Netherlands}
\thanks{$^{11}$Facebook AI Research, Menlo Park, CA, USA}
\thanks{This version of the paper was accepted for publication by IEEE Transactions on Medical Imaging. It has been uploaded to arXiv to comply with IEEE policy. Please see the official version at https://doi.org/10.1109/TMI.2021.3075856.}}

%
%

%

\maketitle

\begin{abstract}
Accelerating MRI scans is one of the principal outstanding problems in the MRI research community.
Towards this goal, we hosted the second fastMRI competition targeted towards reconstructing MR images with subsampled k-space data.
We provided participants with data from 7,299 clinical brain scans (de-identified via a HIPAA-compliant procedure by NYU Langone Health), holding back the fully-sampled data from 894 of these scans for challenge evaluation purposes.
In contrast to the 2019 challenge, we focused our radiologist evaluations on pathological assessment in brain images.
We also debuted a new Transfer track that required participants to submit models evaluated on MRI scanners from outside the training set.
We received 19 submissions from eight different groups.
Results showed one team scoring best in both SSIM scores and qualitative radiologist evaluations.
We also performed analysis on alternative metrics to mitigate the effects of background noise and collected feedback from the participants to inform future challenges.
Lastly, we identify common failure modes across the submissions, highlighting areas of need for future research in the MRI reconstruction community.
\end{abstract}

\begin{IEEEkeywords}
Challenge, Public Data Set, MR Image Reconstruction, Machine Learning, Parallel Imaging, Compressed Sensing, Fast Imaging, Optimization
\end{IEEEkeywords}

\IEEEpeerreviewmaketitle

\section{Introduction}
Due to advances in algorithms, software platforms~\cite{chetlur2014cudnn,abadi2016tensorflow,paszke2019pytorch} and compute hardware, over the last five years there has been a surge of research of MR image reconstruction methods based on machine learning~\cite{sun2016deep,hammernik2018learning,schlemper2017deep,yang2017dagan,eo2018kiki,aggarwal2018modl,zhu2018image,Knoll2019,Knoll2020IEEE,yaman2020self,sriram2020end}.
Traditionally, research in MR image reconstruction methods has been conducted on small data sets collected by individual research groups with direct access to MR scanner hardware and research agreements with the scanner vendors.
Data set collection is difficult and expensive, with many research groups lacking the organizational infrastructure to collect data at the scale necessary for machine learning research.
Furthermore, data sets collected by individual groups are often not shared publicly for a variety of reasons.
As a result, research groups lacking large-scale data collection infrastructure face substantial barriers to reproducing results and making comparisons to existing methods in the literature.

Such challenges have been seen before. In the field of computer vision, the basic principles of convolutional neural networks (CNNs) were proposed as early as 1980~\cite{fukushima1980neocognitron} and became well-established for character recognition by 1998~\cite{lecun1998gradient}.
Following Nvidia's release of CUDA in 2007, independent research groups began to use GPUs to train larger and deeper networks~\cite{raina2009large,cirecsan2010deep}.
Nonetheless, universal acceptance of the utility of CNNs did not occur until the debut of the large-scale ImageNet data set and competition~\cite{deng2009imagenet}.
The introduction of ImageNet allowed direct cross-group comparison using this well-recognized data set of a size beyond what most groups could attain individually.
In 2012 a CNN-based model~\cite{krizhevsky2012imagenet} out-performed all non-CNN models, spurring a flurry of state-of-the-art results for image recognition~\cite{simonyan2014very,he2016deep,szegedy2016rethinking,huang2017densely}.

Since 2018, the fastMRI project has attempted to advance community-based scientific synergy in MRI by building on two pillars.
The first consists of the release of a large data set of raw k-space and DICOM images~\cite{zbontar2018fastmri,knoll2020fastmri}.
This data set is available to almost any researcher, allowing them to download it, replicate results, and make comparisons.
The second pillar consists of hosting public leaderboards~\cite{zbontar2018fastmri} and open competitions, such as the 2019 fastMRI Reconstruction Challenge on knee data~\cite{knoll2020advancing}.
The dimension of public competitions is not new to the MR community.
Other groups have facilitated challenges around RF pulse design~\cite{grissom2017advancing}, diffusion tractography reconstruction~\cite{schilling2019challenges}, and ISMRM initiatives for reconstruction~\cite{maier2020cg,beauferris2020multichannel}.

The 2020 fastMRI Challenge continues this tradition of open competitions and follows the 2019 challenge with a few key differences.
First, our target anatomy has been changed to focus on images of the brain rather than knee.
Second, for 2020 we updated the radiologist evaluation process, asking radiologists to rate images based on \textit{depiction of pathology} rather than \textit{overall image quality}, emphasizing clinical relevance in competition results.
Lastly, we address a core traditional problem in MR imaging: the capacity of models to generalize across sites and vendors.
We introduce a new competition track: a ``Transfer'' track, where participants were asked to run their models on data from vendors not included in training.
This contrasts with the 2019 challenge, which only included data from a single vendor for both training and evaluation.

\section{Methods}
This challenge focuses on MRI scan acceleration, a topic of interest to the MR imaging community for decades.
MRI scanners acquire collections of Fourier frequency ``lines'', commonly referred to as k-space data.
Due to hardware constraints on how magnetic fields can be manipulated, the rate at which these lines are acquired is fixed, which results in relatively long scan times and has negative implications with regard to image quality, patient discomfort, and accessibility.
The major way to decrease scan acquisition time is to decrease the amount of data acquired.
Sampling theory~\cite{whittaker1915xviii,nyquist1928certain,kotel1933carrying,shannon1949communication} states that a minimum number of lines are required for image reconstruction.
This minimum requirement can be circumvented by incorporating other techniques such as parallel imaging~\cite{sodickson1997simultaneous,pruessmann1999sense,griswold2002generalized} and compressed sensing~\cite{lustig2008compressed}.
More recently, machine learning methods have demonstrated further accelerations over parallel imaging and compressed sensing methods.

To promote the advancement of methods for accelerated MRI, we organized a public challenge.
We applied retrospective downsampling to fully-sampled MRIs and provided the downsampled data to challenge participants.
Challenge participants ran their models on the downsampled data and submitted it to the competition website at \url{https://fastmri.org}, where we quantitatively evaluated it using the fully-sampled data as gold standards.
We selected six cases for each of the top three teams in each track of the challenge (three tracks total) and presented the cases to a group of six radiologists for qualitative evaluation.
The challenge winner was selected based on the best depiction of pathology compared to the ground truth as judged by radiologists.

At a high level we describe the principles of our 2020 challenge as follows. Using knowledge we gained through the 2019 challenge, we identified a few key alterations for 2020. These include:
\begin{itemize}
    \item A new imaging anatomy, the brain, the most commonly-imaged organ using MRI.
    \item A focus on an evaluation of pathology depiction rather than overall image quality impressions to strengthen the connection between the challenge evaluation and clinical practice.
    \item An emphasis on generalization with the introduction of a new ``Transfer'' track where participants were asked to run their models on multi-vendor data.
    \item We removed the single-coil track and moved to a pure multi-coil challenge to increase the clinical relevance of the submitted models.
    \item Due to easier practical implementation and removal of the single-coil track, we used pseudo-equispaced subsampling masks (i.e., equispaced masks with a modification for achieving exact 4X/8X sampling rates) rather than random. This follows more closely sampling patterns (and relaxation effects) that are used for parallel imaging in vendor sequences, facilitating easier clinical deployment.
    We maintained the fully-sampled center due to its utility for autocalibrating parallel imaging methods~\cite{sodickson1997simultaneous,griswold2002generalized,uecker2014espirit} and compressed sensing~\cite{lustig2008compressed}.
    \item In the 2019 challenge our baseline model was a U-Net~\cite{ronneberger2015u}; however, winning models~\cite{pezzotti2020adaptive,wang2019pyramid,putzky2019rim} of the 2019 challenge were variational network/cascading models~\cite{knoll2020advancing}. For the 2020 challenge, we provided a much stronger baseline model based on an End-to-End Variational Network~\cite{sriram2020end}.
\end{itemize}

We kept the following principles from the 2019 challenge:
\begin{itemize}
    \item We again used a two-stage evaluation, where a quantitative metric was used to select the top 3 submissions. These finalists were then sent to radiologists to determine the winners. We used the structural similarity index (SSIM)~\cite{wang2004image} as our quantitative image quality index for ranking submissions prior to submission to clinical radiologists~\cite{knoll2020advancing}.
    \item We wanted to maintain realism for a straightforward, 2D imaging setting, and so all of the competition data was once again based on fully-sampled 2D raw k-space data.
    \item For the ground truth reference, we had discussions on alternatives to the root sum-of-squares (RSS) method used for quantitative evaluation in 2019.
    Although there was some consensus on the drawbacks of RSS~\cite{roemer1990nmr,walsh2000adaptive}, there was no consensus on a single best alternative. In the following sections we discuss the impact of this choice further.
\end{itemize}

\subsection{Challenge Tracks}
In the 2019 challenge we included three submission tracks: multicoil with four times acceleration (Multi-Coil 4X), multicoil with eight times acceleration (Multi-Coil 8X), and single-coil with 4X acceleration (Single-Coil 4X).
Among these tracks, the single-coil track garnered the most engagement, but due to its distance from clinical practice we decided to remove it from the 2020 challenge, replacing it with the Transfer track. For the standard multicoil tracks in the 2019 challenge, we observed that although there were many high-quality submissions at 4X, all of the submissions began missing pathology at 8X acceleration~\cite{knoll2020advancing}.
Since this time, 4X machine learning methods have been validated for clinical interchangeability~\cite{recht2020using}.
This suggests that the current upper limit of 2D machine learning image reconstruction performance remains between 4-fold and 8-fold acceleration rates.
In order to provide participants with both an obtainable target and a ``reach'' goal, we kept the 4-fold and 8-fold tracks for the 2020 challenge.

One frequent feedback on the 2019 challenge was on generalizability: despite the size of the data set, all of the data and results were from studies performed on MRI scanners from a single vendor at a single institution.
To address this, we created the new Transfer track at 4-fold acceleration (Transfer 4X).
For the Transfer track, participants were asked to run their models on data from vendors outside the main fastMRI data set.
There was a caveat: we also restricted participants in the Transfer track to train their models only using available fastMRI data to ensure evaluation of transfer capability.
At the time of the 2020 challenge announcement, we stated that these data would come ``from another vendor'' but did not specify further. 
At the challenge launch time, we revealed that the challenge data for this track was a mix of data from GE and Philips, providing additional difficulty for participants.
As a result, submissions in the Transfer track exhibited wide deviations in performance depending on vendor.

\subsection{Data Set}
For the 2020 challenge we used brain MRI data.
The neuroimaging subset of the fastMRI data has been described in an updated version of the arXiv paper~\cite{zbontar2018fastmri}, with further information included in the supplemental material for this paper.
It includes 6,970 scans (3,001 at 1.5 T, 3,969 at 3 T) collected at NYU Langone Health on Siemens scanners using T1, T1 post-contrast, T2, and FLAIR acquisitions.
Unlike the knee challenge, this data set exhibits a wide variety of reconstruction matrix sizes.
A summary of the data for the two main track splits is shown in Table \ref{tab:main_track_data}.
Of these 6,970 scans, 565 were withheld for evaluation in the challenge.
In addition to standard HIPAA-compliant anonymization practices, all scans were cropped at the level of the orbital rim, preserving only the top part of the head.
\begin{table}[htb]
 \caption{Summary of the Challenge Data}
  \centering
  \begin{tabular}{lccccc}
    \toprule
    Split          & T1 & T1POST & T2 & FLAIR &  Total \\
    \midrule \midrule
    \multicolumn{6}{@{}l}{\textbf{Siemens/Main Tracks}} \\ 
    train          & 498 & 949 & 2,678 & 344 & 4,469 \\
    val            & 169 & 287 & 815   & 107 & 1,378 \\
    test (4X)      & 33  & 54  & 170   & 24  & 281   \\
    test (8X)      & 32  & 68  & 152   & 25  & 277   \\
    challenge (4X) & 26  & 67  & 192   & 18  & 303   \\
    challenge (8X) & 24  & 65  & 159   & 14  & 262   \\
    \midrule \midrule
    \multicolumn{6}{@{}l}{\textbf{Transfer Track (4X, all challenge)}} \\
    GE         & 22 & 29     & 83 & 77    & 211    \\
    Philips    & 18 & 0      & 50 & 50    & 118   \\
    \bottomrule
  \end{tabular}
  \label{tab:main_track_data}
\end{table}

For the challenge, the 565 scans were augmented further by 329 non-Siemens scans for the Transfer track.
GE data were collected at NYU Langone Health and Philips data were collected on volunteers by clinical partner sites of Philips Healthcare of North America.
Since the Philips data was collected on volunteers, this subsplit had no post-contrast imaging. 
One difficulty of the Transfer track was the fact that the GE data did not contain frequency oversampling.
The lack of frequency oversampling was due to automatic removal during the analog-to-digital conversion process on the GE scanner.

In total the 2020 challenge had 6,405 scans available for training and validation (train, val, test) and there were 894 total scans for evaluation in the final challenge phase.
This marked a substantial increase in scale from the 2019 challenge.
For reference, the multicoil data from the 2019 challenge on knee data had 1,290 scans for training and validation (train, val, test) and 104 scans for the challenge, so the data for training increased by roughly 5-fold and the data for challenge evaluation increased by roughly 8-fold.

Participants were restricted to only use this data set for training the weights of their models. We did permit participants to use their own data as validation data, but not for backpropagating gradients.

\subsection{Evaluation Process}
Submissions were processed via \url{https://fastmri.org}. This site maintains a submission system for both the challenge and the public test leaderboard. Currently, there are no plans to make the challenge split of the data available in order to maintain the integrity of the results, but research groups may submit to the public leaderboard using the test set via the website at any time.

After submission, evaluation followed a two-stage process of comparisons to the fully-sampled ``ground truth'' images.
For the ground truth images, we followed the previous convention~\cite{knoll2020advancing} to use root sum-of-squares images.
The advantage of this approach is that it does not bias to any one method for coil sensitivity estimation.
There are some drawbacks to RSS, including 1) discarding of the phase information and 2) RSS images can have substantial noise in the background. The phase is not typically used for anatomical evaluation. The issue with noise is more fundamental as it is treated as ground truth in our quantitative evaluation, and any deviations from it influence our ranking.
This is counterbalanced by using radiologist evaluation for declaring the challenge winner.
In planning for the challenge, we were unable to build consensus on an alternative ground truth calculation technique, but this topic could be re-examined in future challenges.
For the quantitative evaluation metric, we chose to use SSIM~\cite{wang2004image}, with a script showing the script used for evaluation in the fastMRI repository at \url{https://github.com/facebookresearch/fastMRI}. SSIM has several parameters. We investigated adjusting these parameters prior to challenge launch, but found that they generally did not alter the ranking of methods evaluated in our quality control phase, and so as a result we used the default parameters in scikit-image~\cite{van2014scikit}.

For the qualitative assessment phase, a board-certified neuroradiologist selected six (two T1 post-contrast, two T2, and two FLAIR) cases from the challenge data set in each of the three tracks.
Cases were specifically selected to represent a broad range of neuroimaging pathologies from intracranial tumors and strokes to normal and age-related changes.
The selection process favored cases with more subtle pathologies for the 4X track and more obvious pathologies for the 8X track with the objective that this might yield better granularity for separating methods in the 4X track.
Selected cases included both intraaxial and extraaxial tumors, strokes, microvascular ischemia, white matter lesions, edema, surgical cavities, as well as postsurgical changes and hardware including craniotomies and ventricular shunts.
The Philips data set was constructed from images of volunteers. Therefore small age-related imaging changes were used for ranking in place of pathology.

Six radiologists with 9-16 years of experience (two of whom are radiology division chiefs) were asked to evaluate the 18 selected image volumes for each team, basing their overall ranking on the quality of the depiction of the pathology using the ground truth as a reference.
Radiologists came from a wide set of institutions, including the Mayo Clinic, Baylor College of Medicine, NYU Langone Health, the University of Pittsburgh Medical Center, Stanford University, and the University of California, Los Angeles.
None of these institutions had finalist submissions.
All radiologists looked at all images in the selected cases during the qualitative evaluation phase, and results were averaged.
Radiologists were aware of the overarching goals of the challenge but were blinded as to which teams submitted the images.
In addition, we also asked radiologists to score each case in terms of artifacts, sharpness and contrast-to-noise ratio (CNR) using a Likert-type scale.
On the Likert scale, 1 was the best (e.g., no artifacts) and 4 was the worst (e.g., unacceptable artifacts).
A Likert score of 3 would affect diagnostic image quality.

\subsection{Timeline}
The 2020 challenge had the following timeline:
\begin{itemize}
    \item December 19, 2019 - Release of the brain data set and update to the arXiv reference~\cite{zbontar2018fastmri}.
    \item July 9, 2020 - Announcement of the 2020 challenge.
    \item October 1-15, 2020 - Release of the challenge data set and submission window.
    \item October 16-19, 2020 - Calculation of SSIM scores. We selected the top 3 submissions for each track and forwarded them to a panel of radiologists for qualitative evaluation.
    \item October 19-November 1, 2020 - Radiologists evaluated submissions. They were asked to complete a score sheet for each of the 3 tracks which included ranking the submissions for each individual case in terms of overall quality of depiction of pathology.
    \item December 5, 2020 - Publication of the challenge leaderboard with results.
    \item December 12, 2020 - Official announcement of the winners of the three tracks with presentations at the Medical Imaging Meets NeurIPS Workshop.
\end{itemize}

\subsection{Overview of Submission Methodologies}
Here we share a brief description of the methodologies behind each of the submissions that made it to the finalist round for radiologist evaluation. The developers of these submissions are included as co-authors on this paper.

\begin{table}[htb]
    \centering
    \caption{Finalist Model Properties}
    \begin{tabular}{lccccc@{}}
        \toprule
        Team & \# Params & Init. & Coil Meth. & GPUs & Tr. Time \\
        \midrule
        \midrule
        AIRS & 200 M & GRAPPA & ESPIRiT & 4 (V100) & 7 days  \\
        ATB & 21 M & Zero-Filled & U-Net & 8 (TITAN) & 10 days \\ 
        MRR & 16 M & Zero-Filled & ESPIRiT & 1 (V100) & 14 days \\
        Nspin & 155 M & Zero-Filled & U-Net & 1 (V100) & 7 days \\
        Res & 841 k & ESPIRiT & U-Net & 4 (RTX 8000) & 21 days \\
        \bottomrule
    \end{tabular}
    \label{tab:finalist_props}
\end{table}
A brief overview of finalist model properties is shown in Table \ref{tab:finalist_props}. (Team names: ``AIRS'' is AIRS Medical, ``ATB'' is ATB, ``MRR'' is MRRecon, ``Nspin'' is Neurospin, ``Res'' is ResoNNance.)
The number of model parameters ranged from 841,000 in the case of ResoNNance to 200 million in the case of AIRS.
Teams applied GRAPPA~\cite{griswold2002generalized}, ESPIRiT~\cite{uecker2014espirit}, or simple zero-filled initializations.
For coil estimation, teams used either ESPIRiT~\cite{uecker2014espirit} or a simple center-based estimation with U-Net refinement similar to that in the End-to-End Variational Network~\cite{sriram2020end}.
Teams used 1-8 GPUs for training, and training time was between 7 and 21 days.

\paragraph*{AIRS Medical}
The AIRS Medical model used a combination of image- and k-space-domain processing in a fashion analogous (but distinct) from that used in KIKI-Net~\cite{eo2018kiki}.
The model included a data consistency cascade with 4 U-Net stages.
At each convolutional layer of the U-Net~\cite{ronneberger2015u}, the multi-domain processing split the channels into one group that operated in image space and one group that operated in k-space~\cite{eo2018kiki}.
Data consistency was enforced at each layer.
The network was initialized with a GRAPPA estimate~\cite{grissom2017advancing} (preprocessed into reconstruction + residual), and coil sensitivities were estimated using ESPIRiT~\cite{uecker2014espirit}.
Since the sampling pattern was pseudo-equispaced, multiple GRAPPA kernels were used to calculate the GRAPPA images.
AIRS optimized their model using Adam~\cite{Kingma2015AdamAM} over 20 epochs using a batch size of 4 at a learning rate of $10^{-3}$ (decayed to $10^{-4}$ after 15 epochs) with SSIM as the loss function.
Optimization took approximately 7 days on four NVIDIA V100 GPUs.
Code is not publicly available.

\paragraph*{ATB}
The ATB model, called ``Joint-ICNet''~\cite{jun2021joint}, was a 10-iteration unrolled algorithm with CNNs replacing regularization terms in a fashion similar to other recent methods~\cite{hammernik2018learning,sriram2020end,eo2018kiki}.
Joint-ICNet used the U-Net~\cite{ronneberger2015u} at each convolutional layer with the dual-domain processing previously introduced in KIKI-net~\cite{eo2018kiki}.
Joint-ICNet used a zero-filled reconstruction as the initial estimate and and coil sensitivities were calculated by refining a rough central k-space estimate with a U-Net~\cite{ronneberger2015u,sriram2020end}.
ATB optimized Joint-ICNet using Adam~\cite{Kingma2015AdamAM} over 50 epochs using a batch size of one at a learning rate of $10^{-4}$ with SSIM as the loss function.
Training took approximately 10 days using 8 NVIDIA TITAN GPUs.
Code is not publicly available.

\paragraph*{MRRecon}
The MRRecon model, called ``Momentum\_DIHN,'' unrolled the Nesterov momentum algorithm with CNN-based regularization for 12 cascades, 6 pre-cascades, and 0 or 1 post-cascade.
The CNN module is a ``Deep, Iterative, Hierarchical Network'' (DIHN) that extends the Down-Up network~\cite{Yu_2019_CVPR_Workshops} with a hierarchical block design, facilitating memory efficiency over a standard U-Net~\cite{ronneberger2015u}.
Momentum\_DIHN used a zero-filled image as the initial estimate and ESPIRiT for calculating coil sensitivities.
To improve transfer track performance, models with several hyperparameters were ensembled to generate the final images.
MRRecon optimized Momentum\_DIHN using Adam~\cite{Kingma2015AdamAM} for less than 5 epochs using a batch size of one at a learning rate of $10^{-4}$ with a compound L1/MS-SSIM loss function~\cite{pezzotti2020adaptive}.
Training took approximately 14 days on an NVIDIA V100 GPU.
Code is not publicly available.

\paragraph*{Neurospin}
The Neurospin model, called XPDNet~\cite{ramzi2020benchmarking,ramzi2021xpdnet}, is a modular neural network unrolling the Chambolle-Pock algorithm~\cite{chambolle2011first} for 25 iterations.
The model was inspired by the primal-only version of the Primal-Dual net~\cite{adler2018learned}, replacing the vanilla CNN with a multi-level wavelet CNN~\cite{liu2018multi}.
XPDNet used a zero-filled image as the initial estimate and calculated coil sensitivities using a rough central k-space estimate refined by a U-Net~\cite{ronneberger2015u,sriram2020end}.
Neurospin optimized XPDNet using Rectified Adam~\cite{Liu2020OnTV} over 100 epochs using a batch size of one at a learning rate of $10^{-4}$ with a compound L1/MS-SSIM loss function~\cite{pezzotti2020adaptive} (98\% MS-SSIM weight).
Training took approximately 7 days on an NVIDIA V100 GPU.
Code is available at \url{https://github.com/zaccharieramzi/fastmri-reproducible-benchmark}.

\paragraph*{ResoNNance}
ResoNNance used a Recurrent Inference Machine (RIM) that has been previously described~\cite{putzky2017recurrent,lonning2019recurrent,putzky2019rim,beauferris2020multichannel}.
Coil sensitivities were calculating using the center of k-spaced followed by U-Net refinement~\cite{ronneberger2015u}.
RIM used ESPIRiT as the model input calculated from the BART toolbox~\cite{uecker2015berkeley}.
ResoNNance optimized RIM using Adam~\cite{Kingma2015AdamAM} over 90 epochs using a batch size of one at a learning rate of $10^{-3}$ with an SSIM loss function.
Separate models were trained for every field strength (1.5 T, 3 T) and contrast (FLAIR, T1/T1PRE/T1POST, and T2).
Code for the RIM, data loaders, and documentation can be found through the DIRECT repository at \url{https://github.com/directgroup/direct}.

\section{Results}

\subsection{Submission Overview}
For the 2020 challenge we received a total of 19 submissions from eight different groups. Seven groups submitted to the Multi-Coil 4X and Multi-Coil 8X tracks. One of these groups chose not to submit to the Transfer track, while an eighth group submitted only to the Transfer track. As previously, we encourage all submitting groups to publish papers and code used to generate their results.

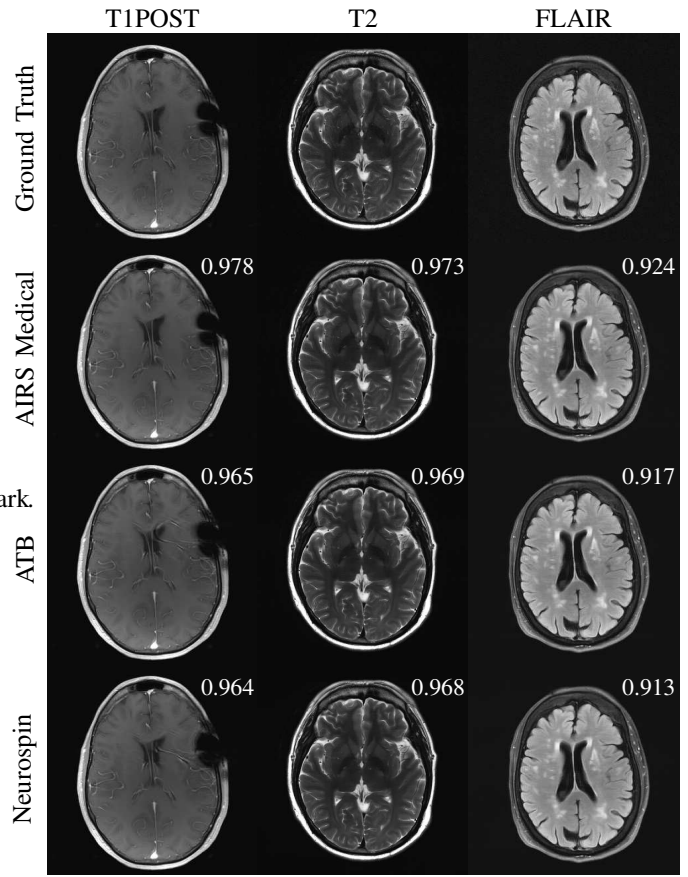
\begin{figure}[htb]
    \centering
    \begin{tikzpicture}
        \draw (0, 1.6cm) node[inner sep=0] (gtlabel) {\raisebox{0in}{\rotatebox[origin=t]{90}{\mbox{}}}};
        \draw (1.7cm, 1.6cm) node[inner sep=0] (gtlabel) {T1POST};
        \draw (4.5cm, 1.6cm) node[inner sep=0] (gtlabel) {T2};
        \draw (7.3cm, 1.6cm) node[inner sep=0] (gtlabel) {FLAIR};
        \draw (0, 0.0cm) node[inner sep=0] (gtlabel) {\raisebox{0in}{\rotatebox[origin=t]{90}{Ground Truth}}};
        \draw (1.7cm, 0.0cm) node[inner sep=0] {\scalebox{-1}[1]{\includegraphics[height=2.8cm]{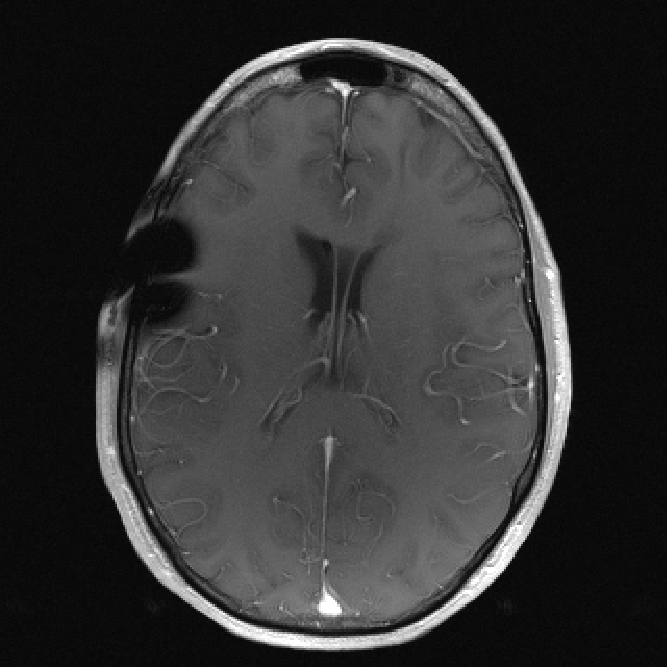}}};
        \draw (4.5cm, 0.0cm) node[inner sep=0] {\scalebox{-1}[1]{\includegraphics[height=2.8cm]{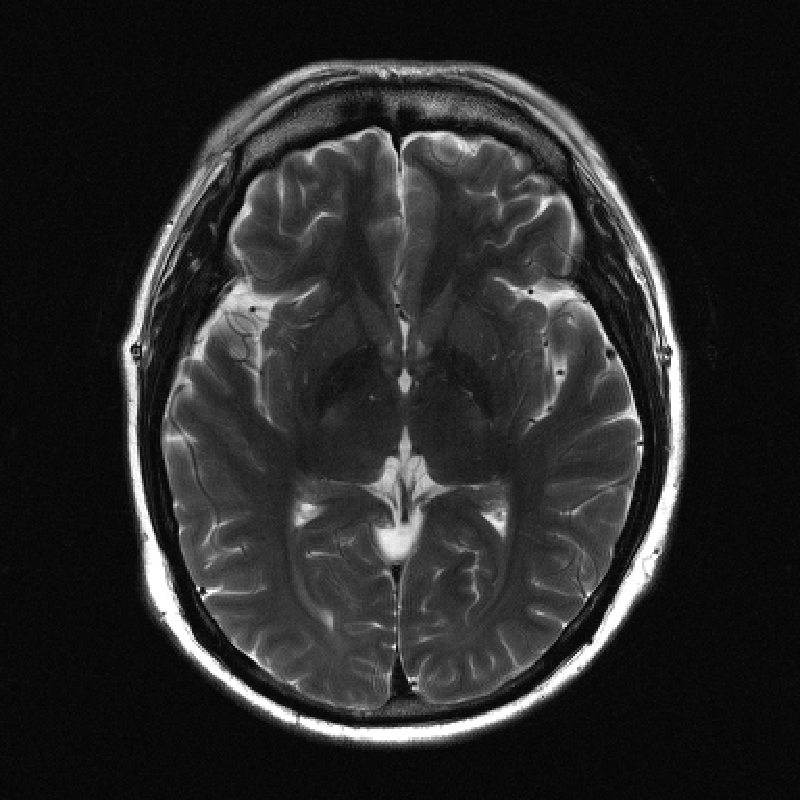}}};
        \draw (7.3cm, 0.0cm) node[inner sep=0] {\scalebox{-1}[1]{\includegraphics[height=2.8cm]{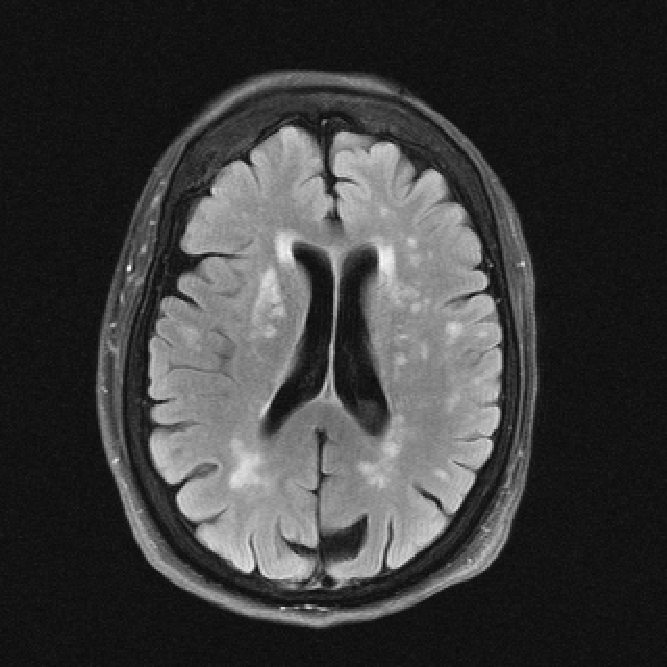}}};
        \draw (0, -2.8cm) node[inner sep=0] (gtlabel) {\raisebox{0in}{\rotatebox[origin=t]{90}{AIRS Medical}}};
        \draw (1.7cm, -2.8cm) node[inner sep=0] {\scalebox{-1}[1]{\includegraphics[height=2.8cm]{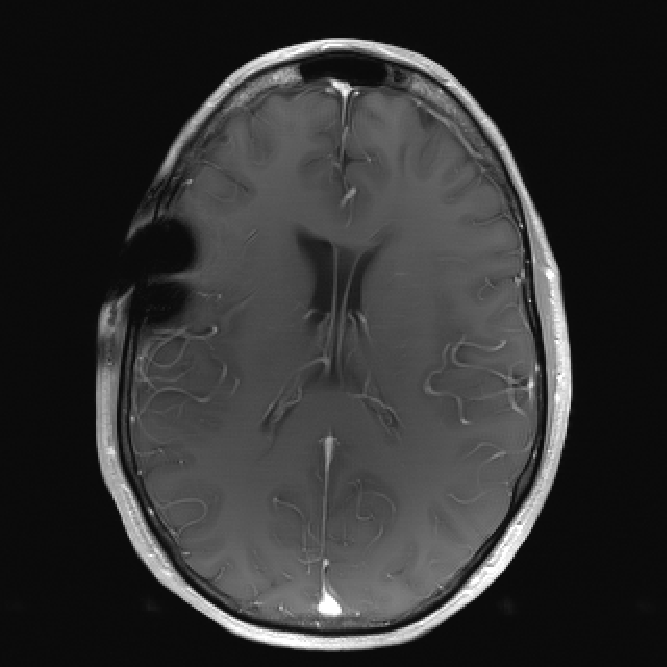}}};
        \draw (2.70cm, -1.70cm) node[inner sep=0] {\small \color{white}0.978};
        \draw (4.5cm, -2.8cm) node[inner sep=0] {\scalebox{-1}[1]{\includegraphics[height=2.8cm]{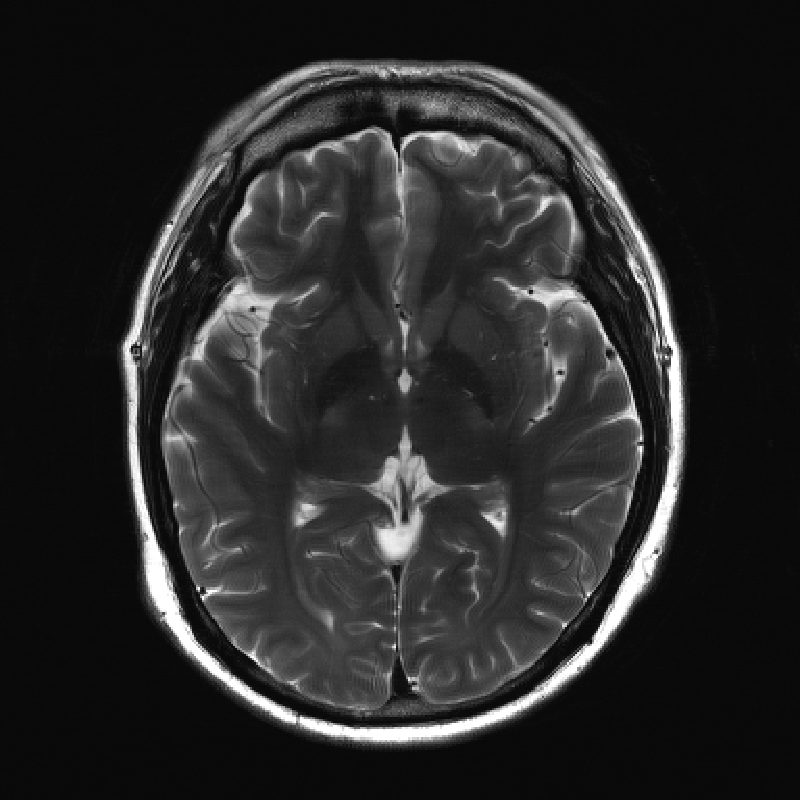}}};
        \draw (5.50cm, -1.70cm) node[inner sep=0] {\small \color{white}0.973};
        \draw (7.3cm, -2.8cm) node[inner sep=0] {\scalebox{-1}[1]{\includegraphics[height=2.8cm]{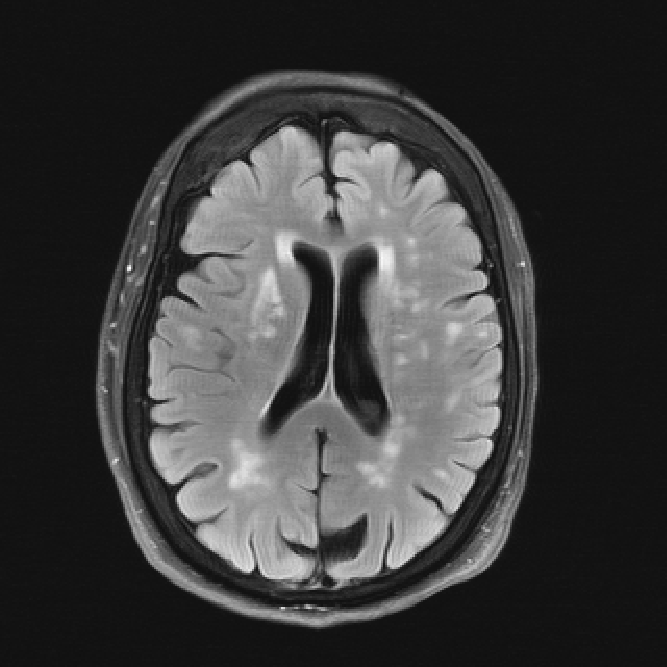}}};
        \draw (8.30cm, -1.70cm) node[inner sep=0] {\small \color{white}0.924};
        \draw (0, -5.6cm) node[inner sep=0] (gtlabel) {\raisebox{0in}{\rotatebox[origin=t]{90}{ATB}}};
        \draw (1.7cm, -5.6cm) node[inner sep=0] {\scalebox{-1}[1]{\includegraphics[height=2.8cm]{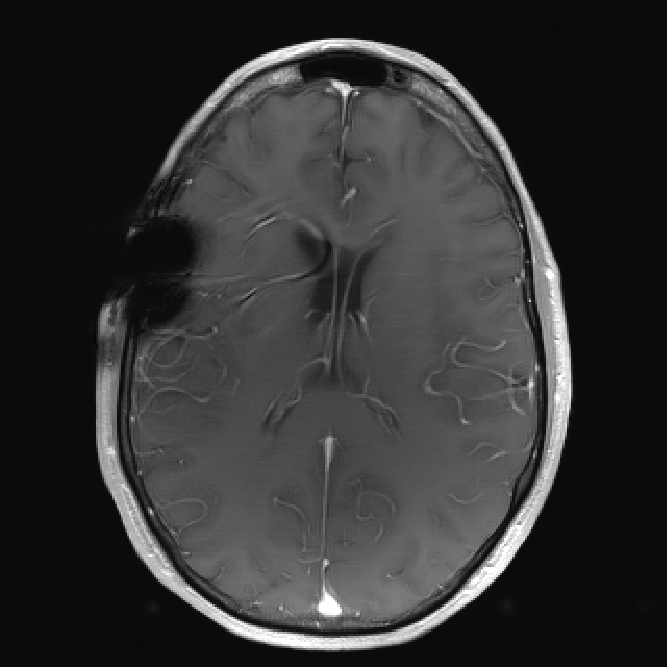}}};
        \draw (2.70cm, -4.50cm) node[inner sep=0] {\small \color{white}0.965};
        \draw (4.5cm, -5.6cm) node[inner sep=0] {\scalebox{-1}[1]{\includegraphics[height=2.8cm]{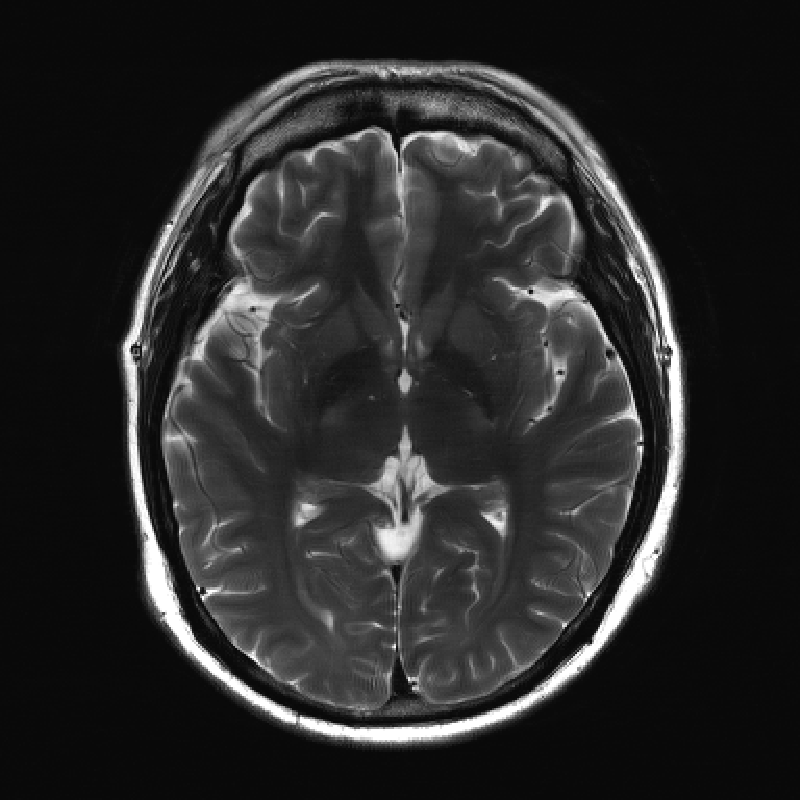}}};
        \draw (5.50cm, -4.50cm) node[inner sep=0] {\small \color{white}0.969};
        \draw (7.3cm, -5.6cm) node[inner sep=0] {\scalebox{-1}[1]{\includegraphics[height=2.8cm]{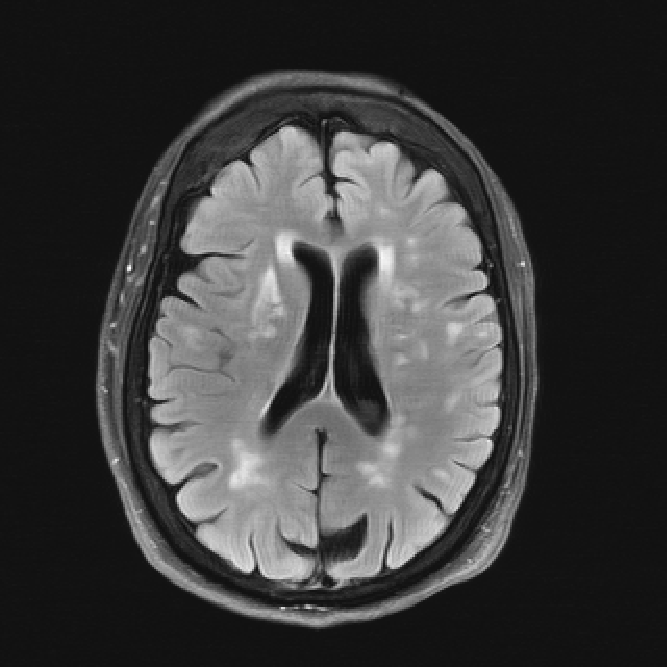}}};
        \draw (8.30cm, -4.50cm) node[inner sep=0] {\small \color{white}0.917};
        \draw (0, -8.399999999999999cm) node[inner sep=0] (gtlabel) {\raisebox{0in}{\rotatebox[origin=t]{90}{Neurospin}}};
        \draw (1.7cm, -8.399999999999999cm) node[inner sep=0] {\scalebox{-1}[1]{\includegraphics[height=2.8cm]{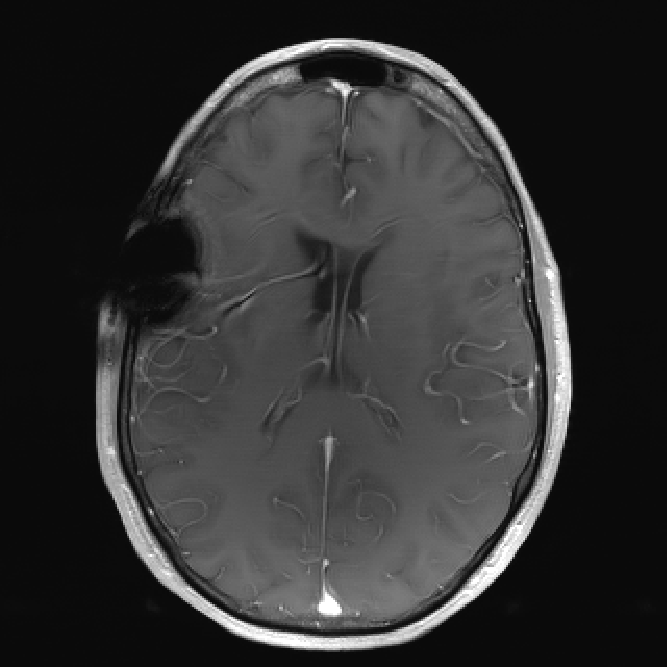}}};
        \draw (2.70cm, -7.30cm) node[inner sep=0] {\small \color{white}0.964};
        \draw (4.5cm, -8.399999999999999cm) node[inner sep=0] {\scalebox{-1}[1]{\includegraphics[height=2.8cm]{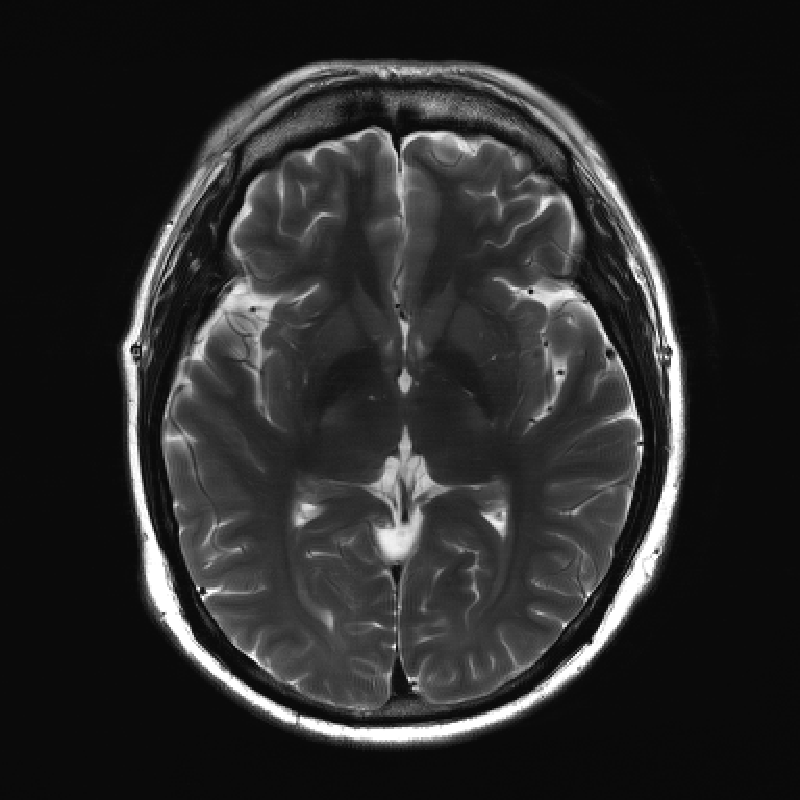}}};
        \draw (5.50cm, -7.30cm) node[inner sep=0] {\small \color{white}0.968};
        \draw (7.3cm, -8.399999999999999cm) node[inner sep=0] {\scalebox{-1}[1]{\includegraphics[height=2.8cm]{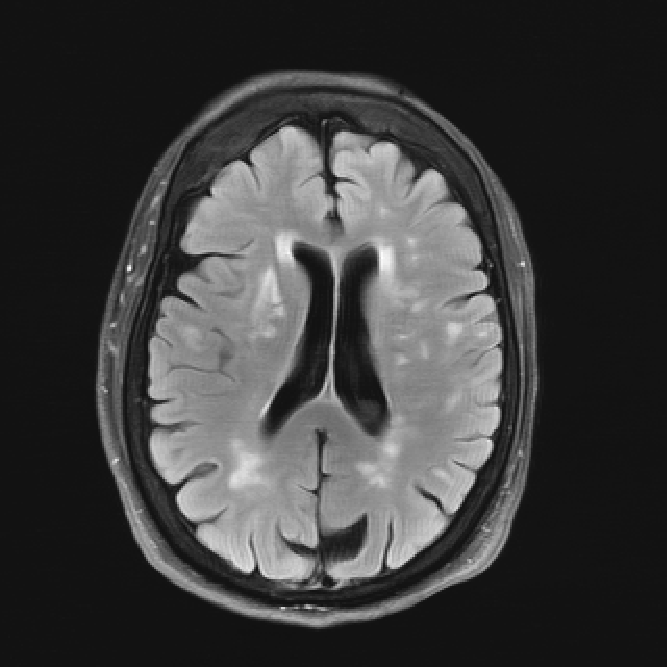}}};
        \draw (8.30cm, -7.30cm) node[inner sep=0] {\small \color{white}0.913};
    \end{tikzpicture}
    \caption{Examples of 4X submissions evaluated by radiologists with slice-level SSIM scores. All methods reasonably reconstructed T2 and FLAIR images. The ATB and Neurospin methods struggled with a susceptibility region, exaggerating the focus of susceptibility and introducing a few false vessels between the susceptibility and the lateral ventricular wall. In other cases, radiologists observed mild smoothing of white matter regions on T1POST images.}
    \label{fig:4x_imgs}
\end{figure}
Figure \ref{fig:4x_imgs} shows an overview of images submitted to the 4X track of the challenge with Siemens data that were forwarded to radiologists.
All three top performing submissions were able to successfully reconstruct the T2 and FLAIR images with minimal artifact presentation.
For some images in this track's evaluation, radiologists had difficulty perceiving substantive differences between the three top performing reconstructions in terms of their overall ability to depict the pathology.
Overall, the results were better on the high signal-to-noise T2 and FLAIR contrasts compared with those on the T1POST.
In the case in Figure \ref{fig:4x_imgs}, the ATB and Neurospin methods struggled with a strong susceptibility effect, introducing false vessels between the susceptibility and the lateral ventricular wall.

\begin{figure}[htb]
    \centering
    \begin{tikzpicture}
        \draw (0, 1.6cm) node[inner sep=0] (gtlabel) {\raisebox{0in}{\rotatebox[origin=t]{90}{\mbox{}}}};
        \draw (1.7cm, 1.6cm) node[inner sep=0] (gtlabel) {T1POST};
        \draw (4.5cm, 1.6cm) node[inner sep=0] (gtlabel) {T2};
        \draw (7.0cm, 1.6cm) node[inner sep=0] (gtlabel) {FLAIR};
        \draw (0, 0.0cm) node[inner sep=0] (gtlabel) {\raisebox{0in}{\rotatebox[origin=t]{90}{Ground Truth}}};
        \draw (1.7cm, 0.0cm) node[inner sep=0] {\scalebox{-1}[1]{\includegraphics[height=2.8cm]{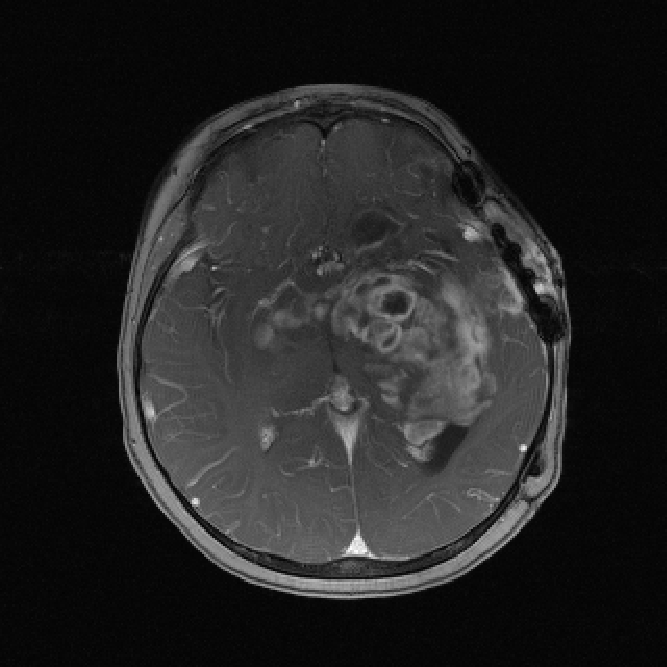}}};
        \draw (4.5cm, 0.0cm) node[inner sep=0] {\scalebox{-1}[1]{\includegraphics[height=2.8cm]{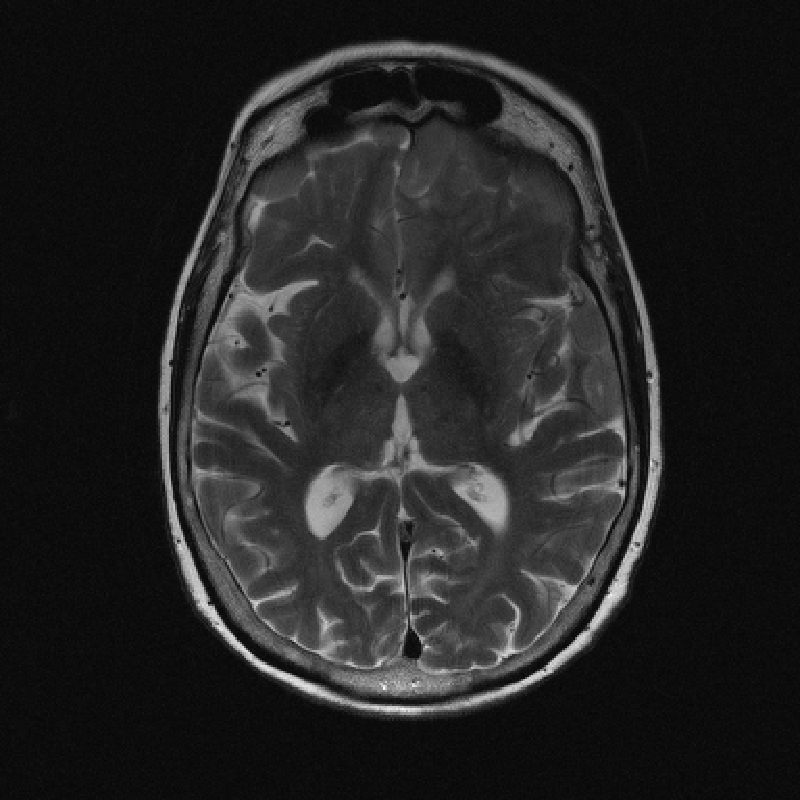}}};
        \draw (7.0cm, 0.0cm) node[inner sep=0] {\scalebox{-1}[1]{\includegraphics[height=2.8cm]{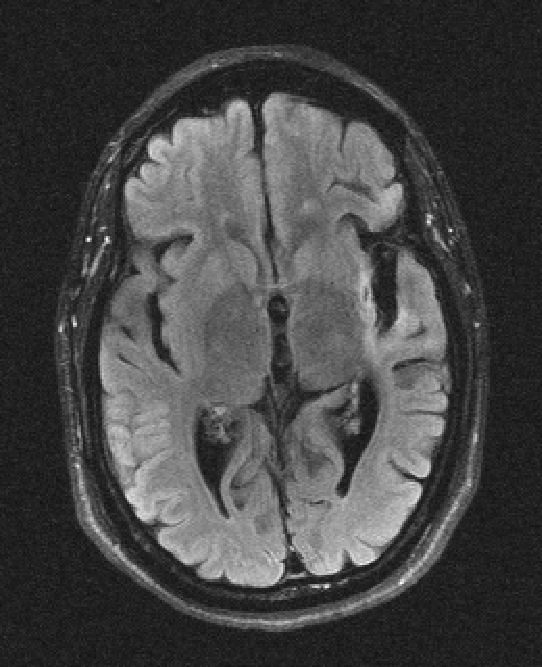}}};
        \draw (0, -2.8cm) node[inner sep=0] (gtlabel) {\raisebox{0in}{\rotatebox[origin=t]{90}{AIRS Medical}}};
        \draw (1.7cm, -2.8cm) node[inner sep=0] {\scalebox{-1}[1]{\includegraphics[height=2.8cm]{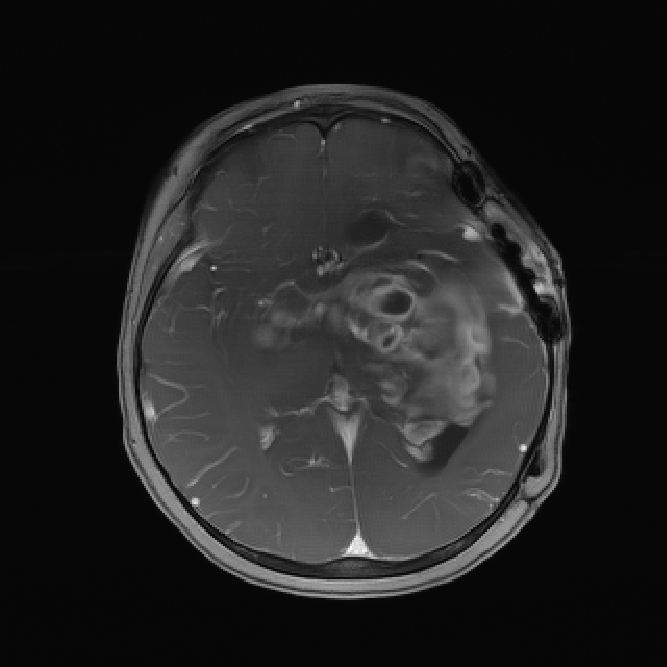}}};
        \draw (2.70cm, -1.70cm) node[inner sep=0] {\small \color{white}0.933};
        \draw (4.5cm, -2.8cm) node[inner sep=0] {\scalebox{-1}[1]{\includegraphics[height=2.8cm]{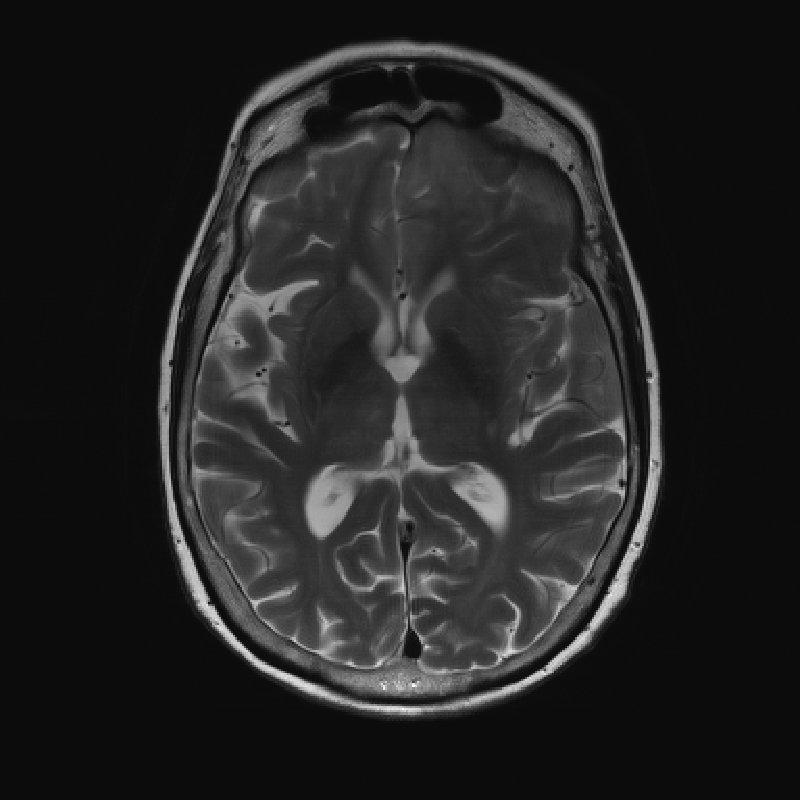}}};
        \draw (5.50cm, -1.70cm) node[inner sep=0] {\small \color{white}0.946};
        \draw (7.0cm, -2.8cm) node[inner sep=0] {\scalebox{-1}[1]{\includegraphics[height=2.8cm]{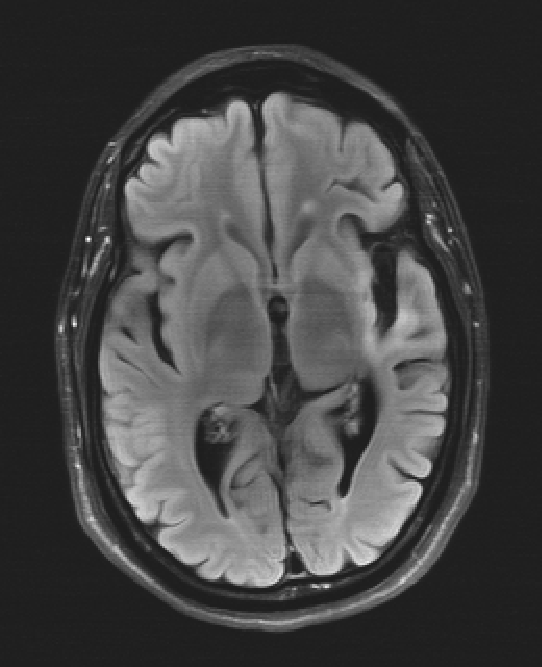}}};
        \draw (7.75cm, -1.70cm) node[inner sep=0] {\small \color{white}0.864};
        \draw (0, -5.6cm) node[inner sep=0] (gtlabel) {\raisebox{0in}{\rotatebox[origin=t]{90}{ATB}}};
        \draw (1.7cm, -5.6cm) node[inner sep=0] {\scalebox{-1}[1]{\includegraphics[height=2.8cm]{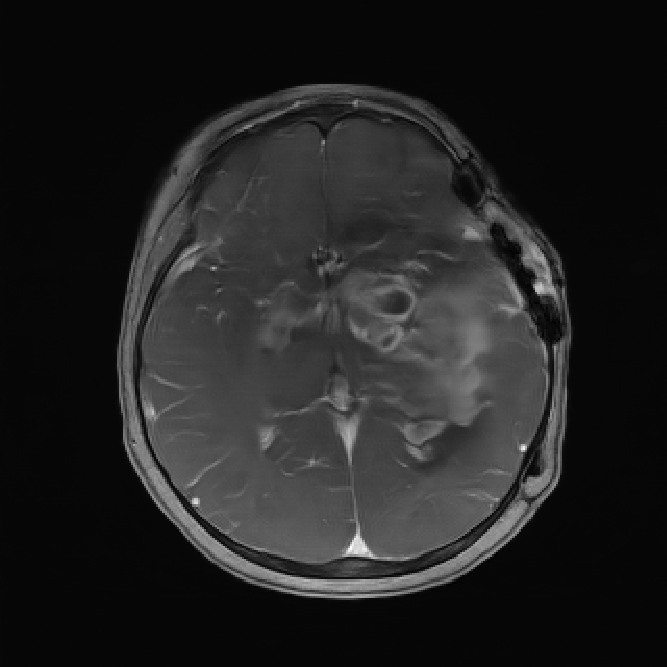}}};
        \draw (2.70cm, -4.50cm) node[inner sep=0] {\small \color{white}0.907};
        \draw (4.5cm, -5.6cm) node[inner sep=0] {\scalebox{-1}[1]{\includegraphics[height=2.8cm]{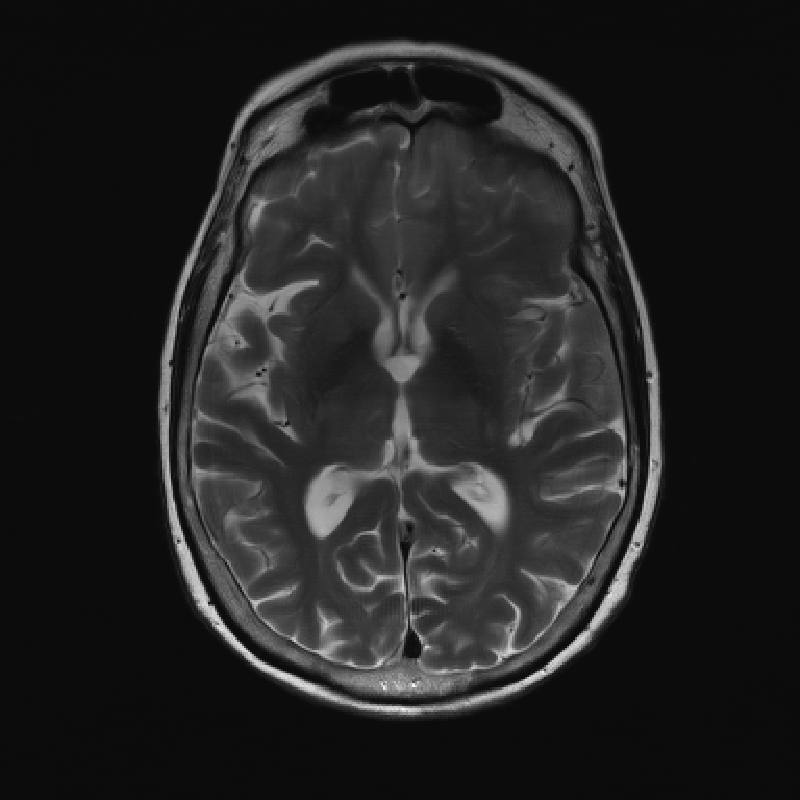}}};
        \draw (5.50cm, -4.50cm) node[inner sep=0] {\small \color{white}0.936};
        \draw (7.0cm, -5.6cm) node[inner sep=0] {\scalebox{-1}[1]{\includegraphics[height=2.8cm]{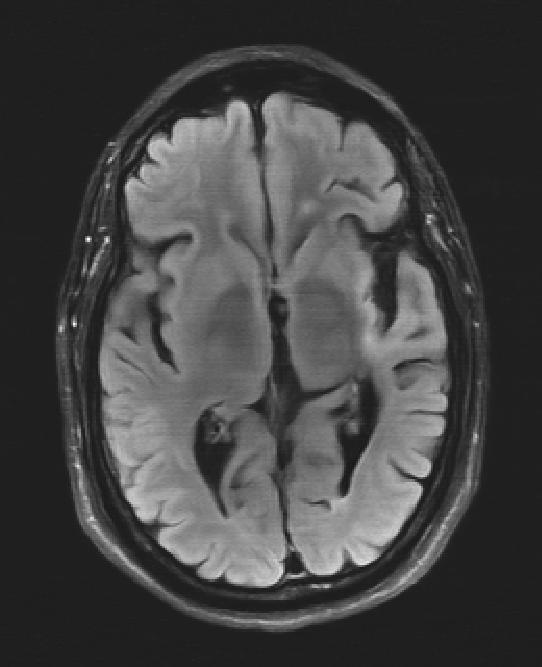}}};
        \draw (7.75cm, -4.50cm) node[inner sep=0] {\small \color{white}0.847};
        \draw (0, -8.399999999999999cm) node[inner sep=0] (gtlabel) {\raisebox{0in}{\rotatebox[origin=t]{90}{Neurospin}}};
        \draw (1.7cm, -8.399999999999999cm) node[inner sep=0] {\scalebox{-1}[1]{\includegraphics[height=2.8cm]{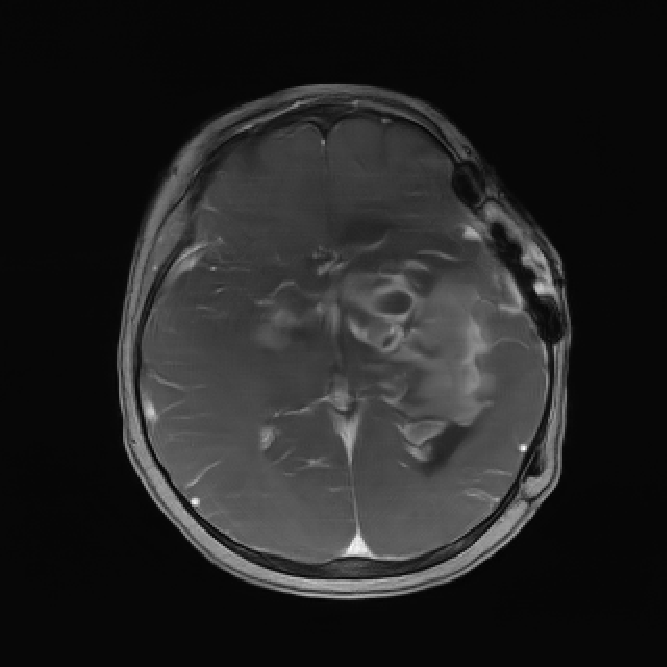}}};
        \draw (2.70cm, -7.30cm) node[inner sep=0] {\small \color{white}0.904};
        \draw (4.5cm, -8.399999999999999cm) node[inner sep=0] {\scalebox{-1}[1]{\includegraphics[height=2.8cm]{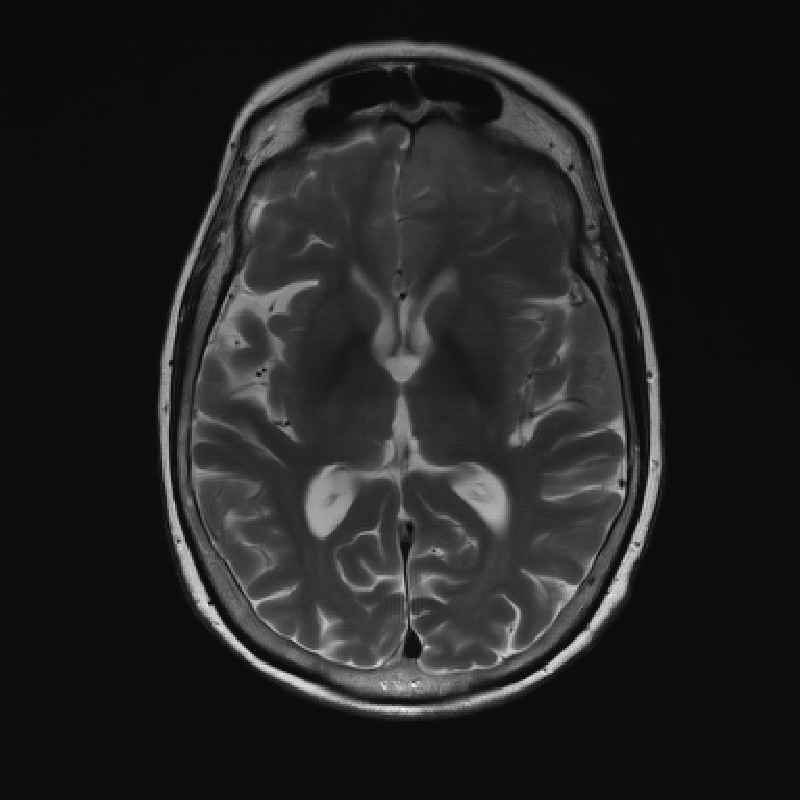}}};
        \draw (5.50cm, -7.30cm) node[inner sep=0] {\small \color{white}0.935};
        \draw (7.0cm, -8.399999999999999cm) node[inner sep=0] {\scalebox{-1}[1]{\includegraphics[height=2.8cm]{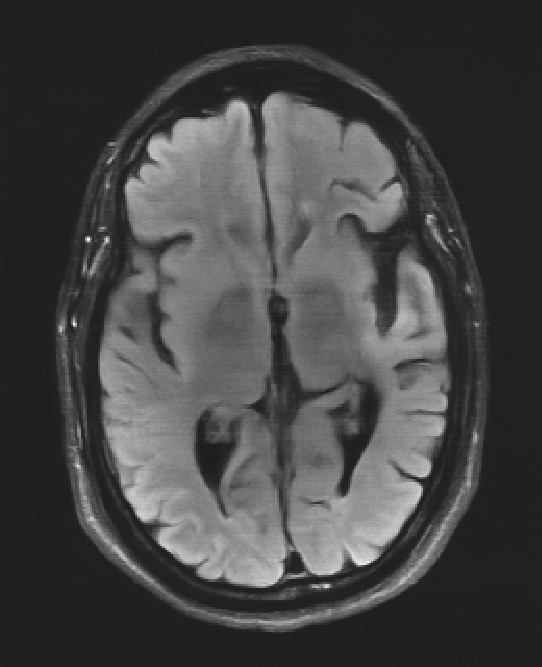}}};
        \draw (7.75cm, -7.30cm) node[inner sep=0] {\small \color{white}0.836};
    \end{tikzpicture}
    \caption{Examples of 8X submissions evaluated by radiologists with slice-level SSIM scores. At this level of acceleration fine details are smoothed and obscured for all contrasts. On T1POST images, AIRS Medical was relatively more successful than ATB and Neurospin in showing fine details of the mass, particularly in its periphery. Noticeable on the FLAIR images are horizontal ``banding'' effects that arise from how neural networks interact with anisotropic sampling patterns.}
    \label{fig:8x_imgs}
\end{figure}
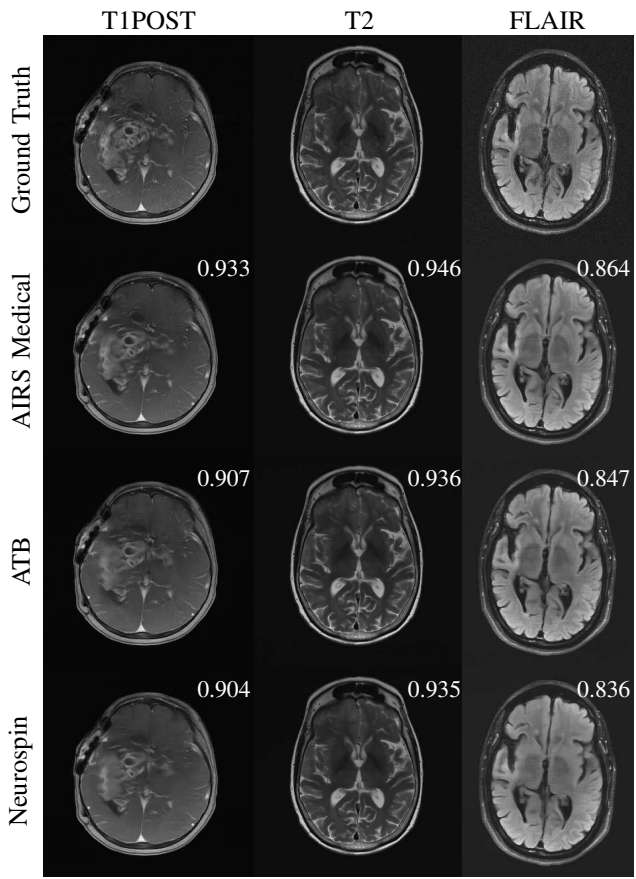
Figure \ref{fig:8x_imgs} shows example images for radiologist evaluation from the 8X track with Siemens data.
In this track, artifacts are seen to be more severe and pronounced.
For some cases radiologists stated that they were hesitant to accept any of the submissions at 8X.
Over-smoothing is readily apparent in T1POST reconstructions from all three of the top performers.
We noticed at this acceleration level that so-called horizontal ``banding'' effects~\cite{defazio2020mri} could be appreciated in the FLAIR images due to the extreme acceleration and the anisotropic sampling pattern.

\begin{figure}[htb]
    \centering
    \begin{tikzpicture}
        \draw (0, 1.6cm) node[inner sep=0] (gtlabel) {\raisebox{0in}{\rotatebox[origin=t]{90}{\mbox{}}}};
        \draw (1.7cm, 1.6cm) node[inner sep=0] (gtlabel) {T1POST};
        \draw (4.5cm, 1.6cm) node[inner sep=0] (gtlabel) {T2};
        \draw (7.3cm, 1.6cm) node[inner sep=0] (gtlabel) {FLAIR};
        \draw (0, 0.0cm) node[inner sep=0] (gtlabel) {\raisebox{0in}{\rotatebox[origin=t]{90}{Ground Truth}}};
        \draw (1.7cm, 0.0cm) node[inner sep=0] {\scalebox{-1}[1]{\includegraphics[height=2.8cm]{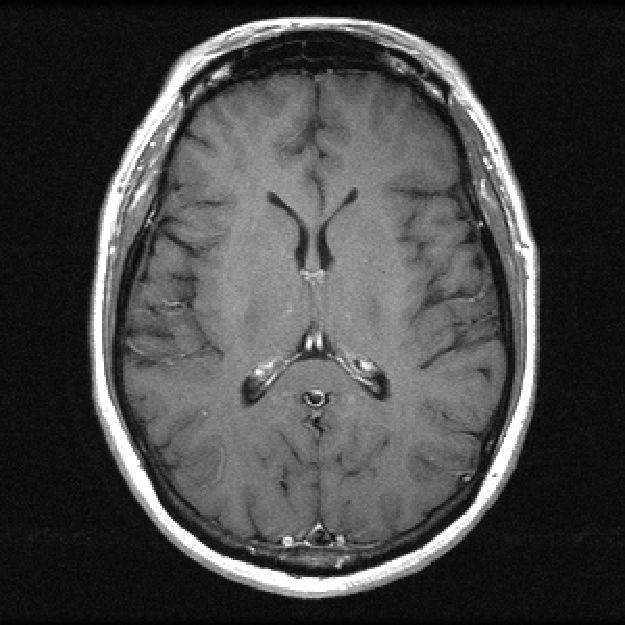}}};
        \draw (4.5cm, 0.0cm) node[inner sep=0] {\scalebox{-1}[1]{\includegraphics[height=2.8cm]{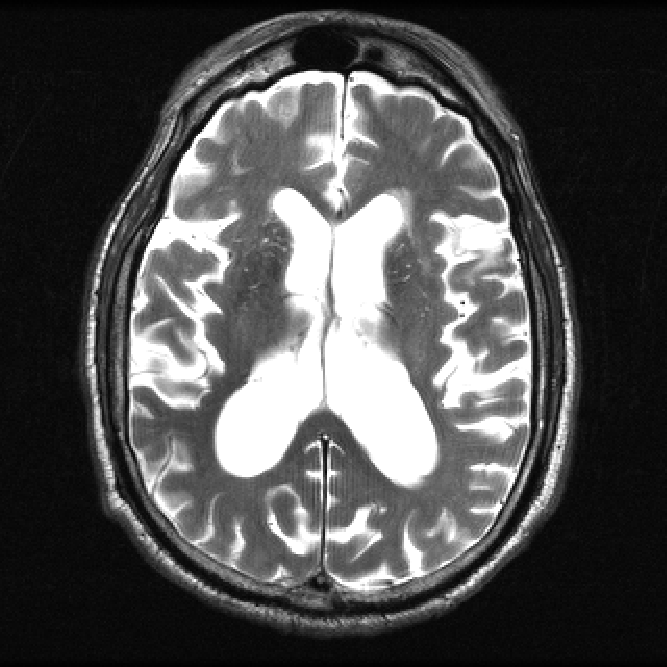}}};
        \draw (7.3cm, 0.0cm) node[inner sep=0] {\scalebox{-1}[1]{\includegraphics[height=2.8cm]{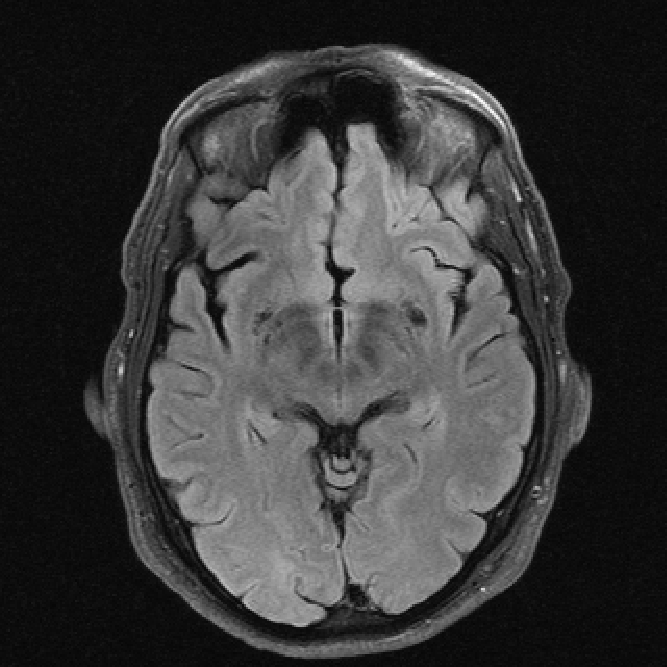}}};
        \draw (0, -2.8cm) node[inner sep=0] (gtlabel) {\raisebox{0in}{\rotatebox[origin=t]{90}{AIRS Medical}}};
        \draw (1.7cm, -2.8cm) node[inner sep=0] {\scalebox{-1}[1]{\includegraphics[height=2.8cm]{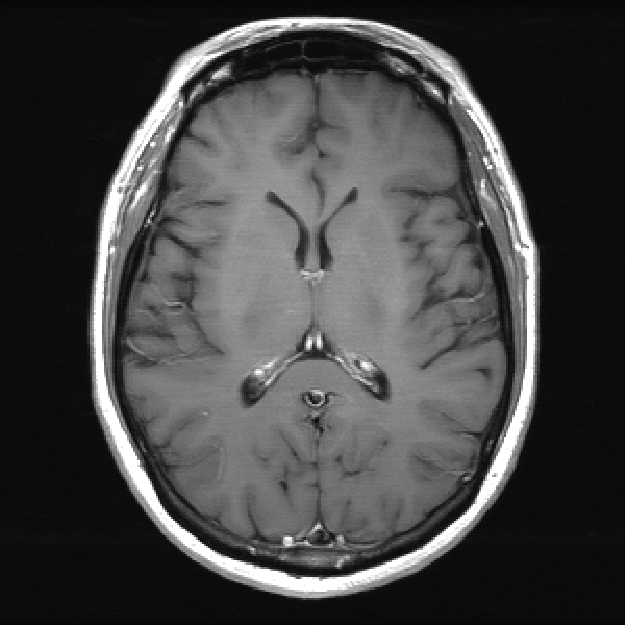}}};
        \draw (2.70cm, -1.65cm) node[inner sep=0] {\small \color{white}0.964};
        \draw (4.5cm, -2.8cm) node[inner sep=0] {\scalebox{-1}[1]{\includegraphics[height=2.8cm]{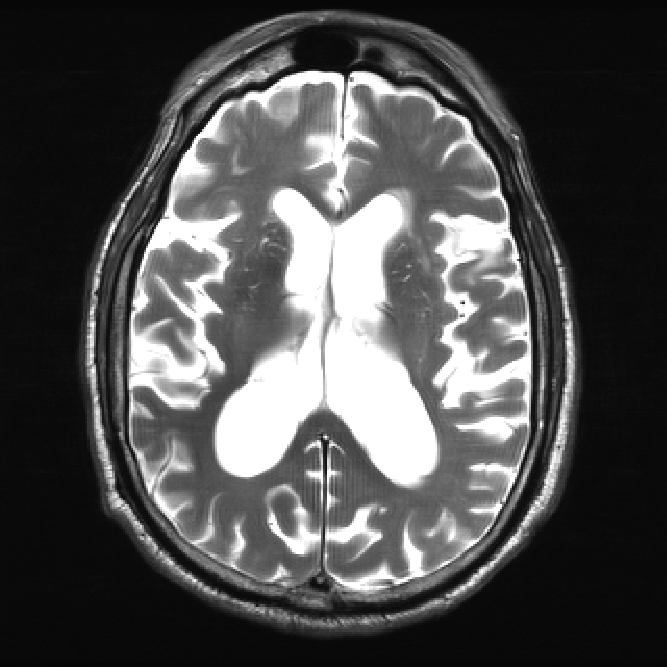}}};
        \draw (5.50cm, -1.65cm) node[inner sep=0] {\small \color{white}0.970};
        \draw (7.3cm, -2.8cm) node[inner sep=0] {\scalebox{-1}[1]{\includegraphics[height=2.8cm]{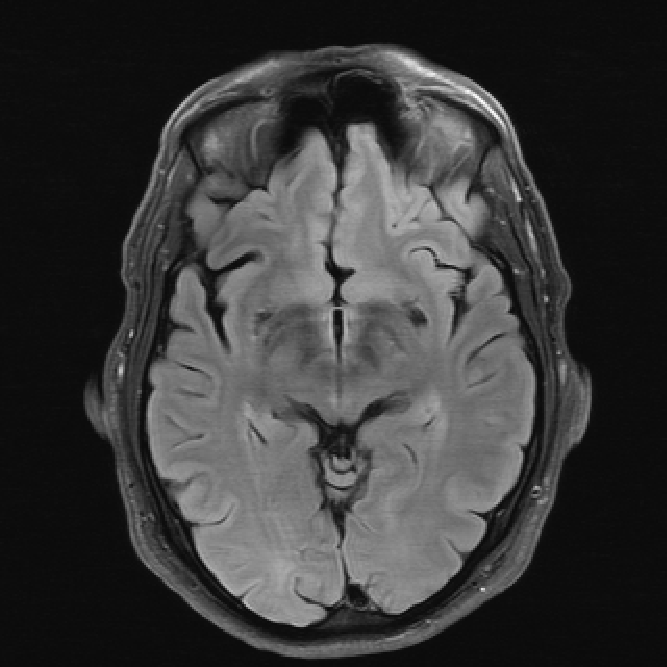}}};
        \draw (8.30cm, -1.65cm) node[inner sep=0] {\small \color{white}0.947};
        \draw (0, -5.6cm) node[inner sep=0] (gtlabel) {\raisebox{0in}{\rotatebox[origin=t]{90}{MRRecon}}};
        \draw (1.7cm, -5.6cm) node[inner sep=0] {\scalebox{-1}[1]{\includegraphics[height=2.8cm]{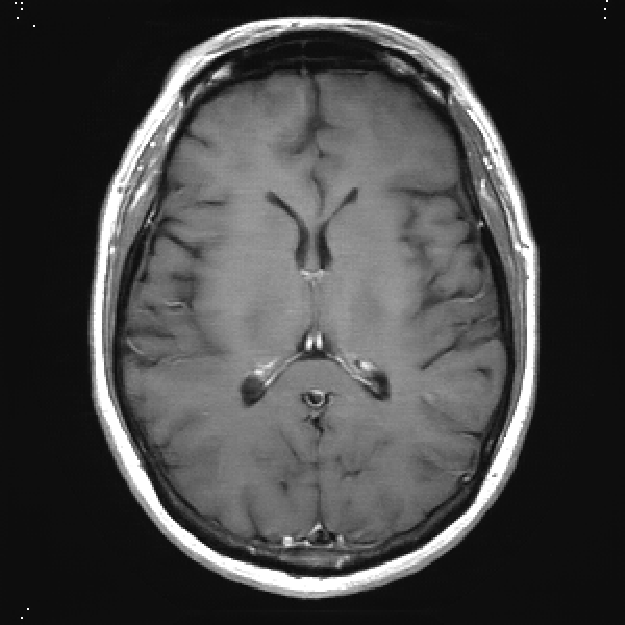}}};
        \draw (2.70cm, -4.45cm) node[inner sep=0] {\small \color{white}0.960};
        \draw (4.5cm, -5.6cm) node[inner sep=0] {\scalebox{-1}[1]{\includegraphics[height=2.8cm]{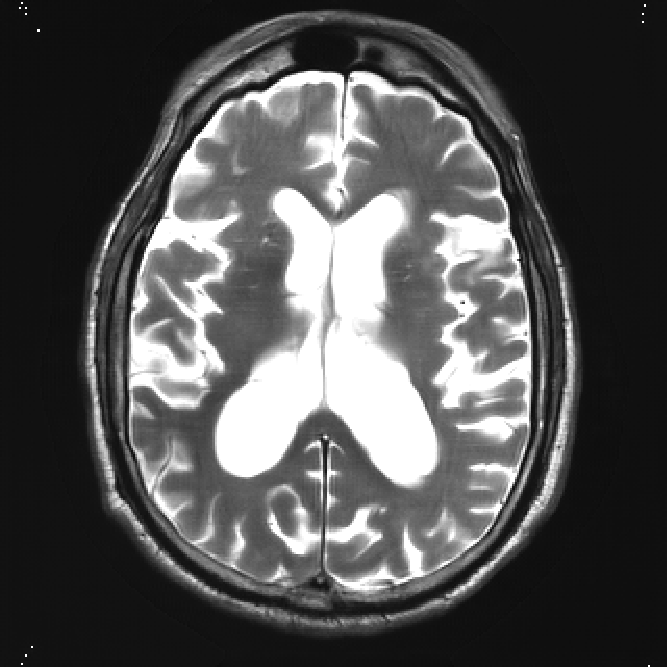}}};
        \draw (5.50cm, -4.45cm) node[inner sep=0] {\small \color{white}0.924};
        \draw (7.3cm, -5.6cm) node[inner sep=0] {\scalebox{-1}[1]{\includegraphics[height=2.8cm]{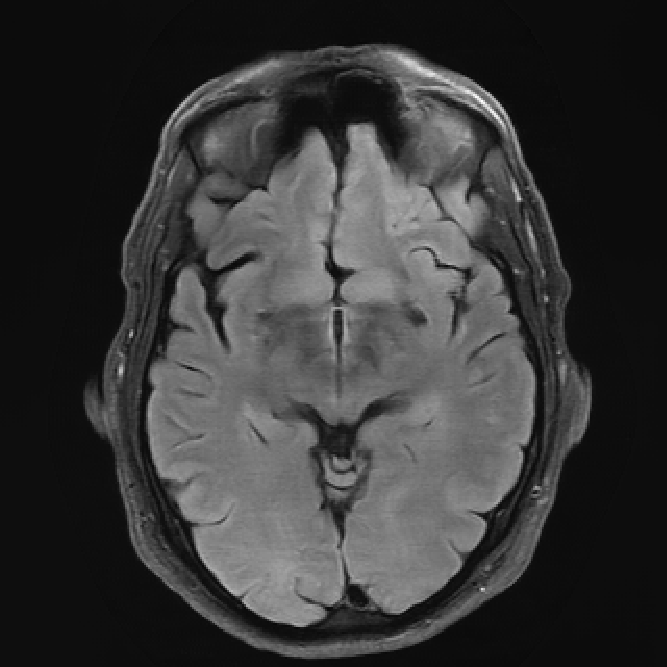}}};
        \draw (8.30cm, -4.45cm) node[inner sep=0] {\small \color{white}0.933};
        \draw (0, -8.399999999999999cm) node[inner sep=0] (gtlabel) {\raisebox{0in}{\rotatebox[origin=t]{90}{ResoNNance}}};
        \draw (1.7cm, -8.399999999999999cm) node[inner sep=0] {\scalebox{-1}[1]{\includegraphics[height=2.8cm]{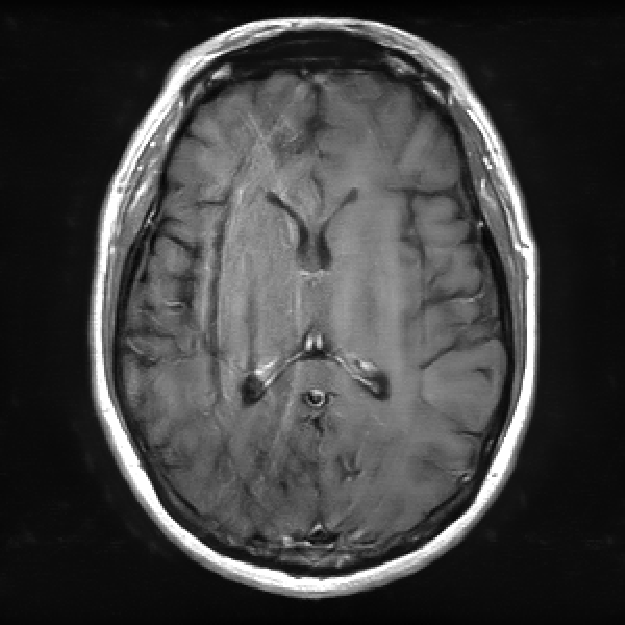}}};
        \draw (2.70cm, -7.25cm) node[inner sep=0] {\small \color{white}0.919};
        \draw (4.5cm, -8.399999999999999cm) node[inner sep=0] {\scalebox{-1}[1]{\includegraphics[height=2.8cm]{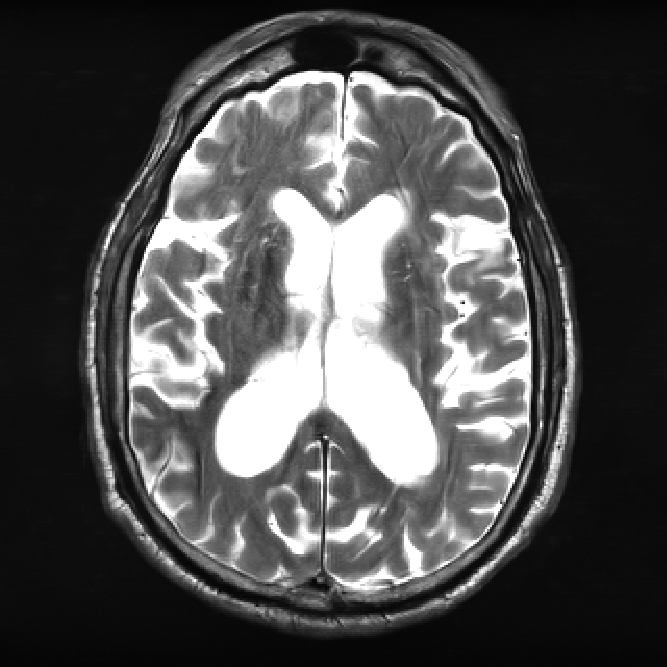}}};
        \draw (5.50cm, -7.25cm) node[inner sep=0] {\small \color{white}0.930};
        \draw (7.3cm, -8.399999999999999cm) node[inner sep=0] {\scalebox{-1}[1]{\includegraphics[height=2.8cm]{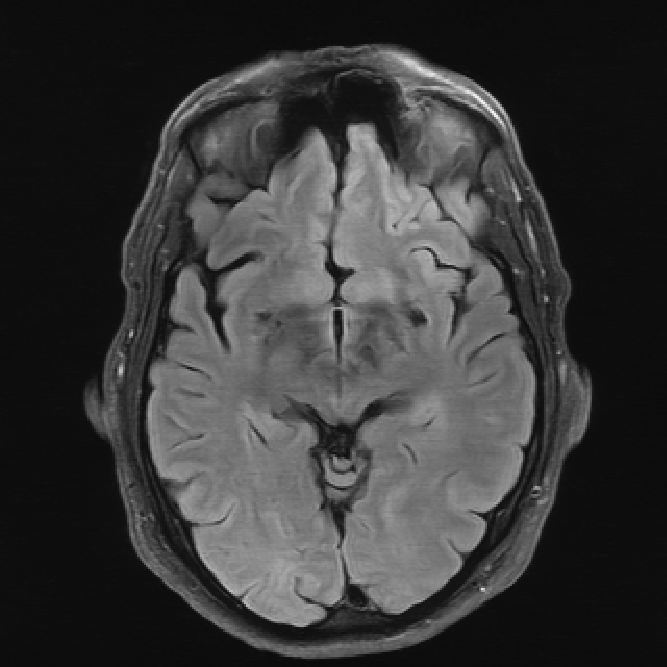}}};
        \draw (8.30cm, -7.25cm) node[inner sep=0] {\small \color{white}0.932};
    \end{tikzpicture}
    \caption{Examples of 4X Transfer submissions evaluated by radiologists with slice-level SSIM scores. The T1POST and T2 examples are from GE scanners, whereas the FLAIR example is from a Philips scanner. All methods introduced blurring to the images. Several methods had trouble adapting to the GE data while performing relatively well on the Philips data, as seen in the form of aliasing artifacts in one of the T1POST images.}
    \label{fig:transfer_imgs}
\end{figure}
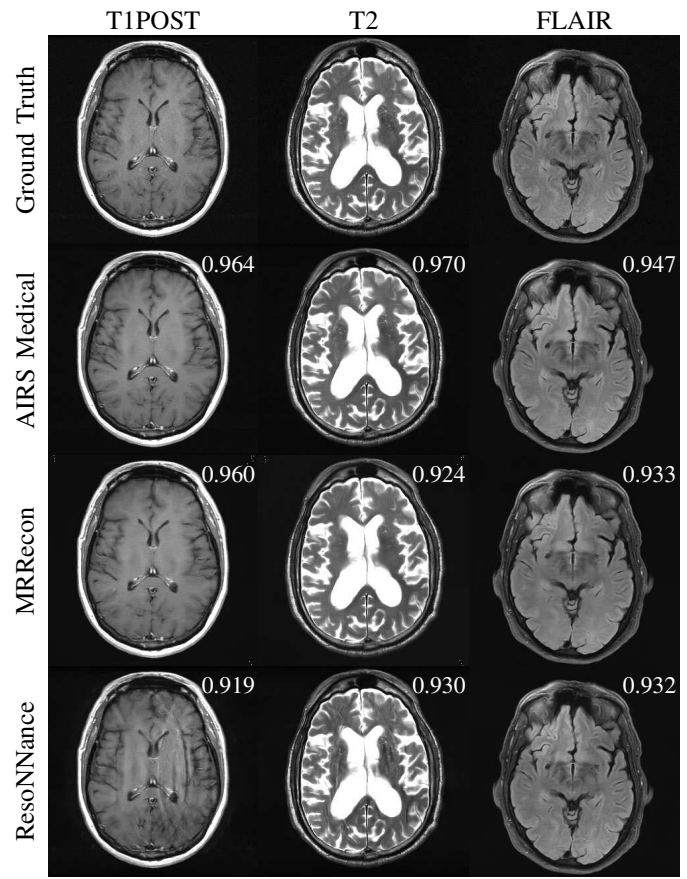
Example images from the 4X Transfer track are shown in Figure \ref{fig:transfer_imgs}.
For this track, we observed the lowest SSIM values (Section \ref{subsec:quantitativeresults}).
Of note, there is a divergence between performance of methods on GE versus Philips data.
This can be seen the image submitted by ResoNNance in Figure \ref{fig:transfer_imgs}, which introduces artifacts in its reconstructions of the GE images (T1POST and T2 in Figure \ref{fig:transfer_imgs}), but less so in its Philips reconstruction (FLAIR in Figure \ref{fig:transfer_imgs}).
Most participant models (trained on Siemens data) were able to reconstruct Philips data with higher fidelity than GE, likely due to the fact that Philips and Siemens followed the same protocol for writing frequency-oversampled data to their raw data files.
An additional factor is that GE uses a T1-based FLAIR, whereas Philips and Siemens use a T2-based FLAIR.

\subsection{Quantitative Results}
\label{subsec:quantitativeresults}
\begin{figure}[htb]
    \centering
    \begin{subfigure}{0.43\textwidth}
        \includegraphics[width=1\linewidth]{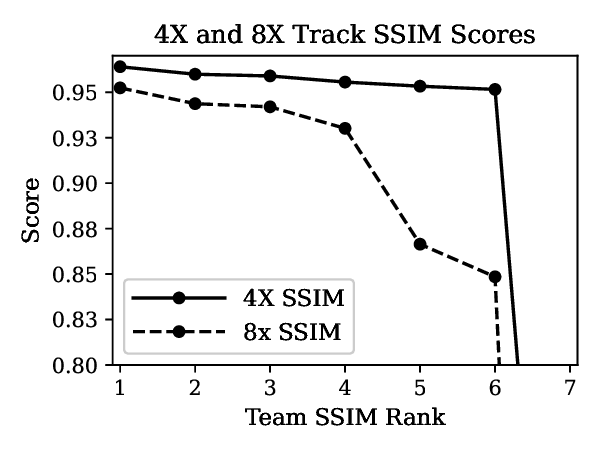}
    \end{subfigure}
    
    \begin{subfigure}{0.43\textwidth}
        \includegraphics[width=1\linewidth]{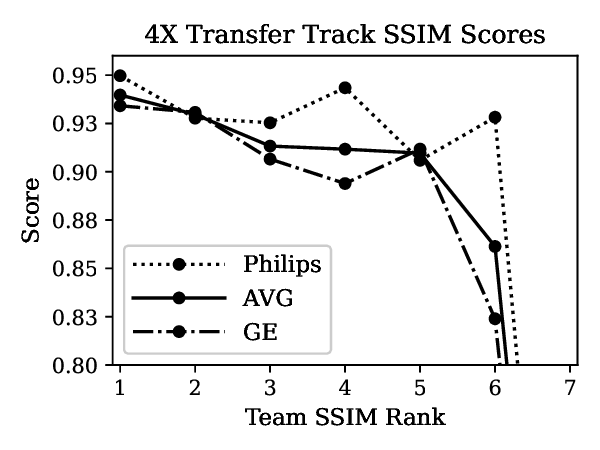}
    \end{subfigure}
    \caption{Summary of SSIM values across contestants. (\textit{top}) Model perfomance for teams submitting to the main 4X and 8X Siemens competition tracks. (\textit{bottom}) Model performance for teams submitting to the Transfer track (combination of GE and Philips data). The ``AVG'' model score for the Transfer track was a simple average across all volumes in the Transfer track.}
    \label{fig:ssim_val_rankings}
\end{figure}

Figure \ref{fig:ssim_val_rankings} shows an overview of SSIM scores across group rankings.
SSIM values were highly clustered in the 4X track, with all top 4 participants scoring between 0.955 and 0.965.
We observed greater variation between submissions in the 8X track, with the top participant scoring 0.952 and the others scoring below 0.944.
The greatest variation occurred in the Transfer track.
Many participants struggled to adapt their models to GE data.
These data did not include frequency oversampling in the raw k-space data, which we have observed can decrease SSIMs for models by as much as 0.1-0.4 if no other adjustments are made.
On the other hand, the Philips data did include frequency oversampling, so adaptation here was more straightforward.

\begin{table*}[htb]
    \centering
    \caption{Summary of Average SSIM Scores with 95\% Confidence Intervals}
    \begin{tabular}{lccccc@{}}
        \toprule
        Team         & AVG   & T1    & T1POST & T2   & FLAIR \\
        \midrule \midrule
        \multicolumn{6}{@{}l}{\textbf{4X Track}}\\
        AIRS Medical & $\textbf{0.964 (0.914, 0.987)}$ & $\textbf{0.967 (0.926, 0.985)}$ & $\textbf{0.969 (0.935, 0.986)}$ & $\textbf{0.965 (0.933, 0.988)}$ & $\textbf{0.930 (0.862, 0.976)}$  \\ 
        ATB & 0.960 (0.908, 0.985) & 0.964 (0.922, 0.983) & 0.965 (0.928, 0.984) & 0.961 (0.926, 0.986) & 0.924 (0.850, 0.971)  \\ 
        Neurospin & 0.959 (0.904, 0.984) & 0.963 (0.920, 0.982) & 0.965 (0.926, 0.984) & 0.960 (0.924, 0.986) & 0.920 (0.844, 0.970)  \\ 
        \midrule \midrule
        \multicolumn{6}{@{}l}{\textbf{8X Track}}\\
        AIRS Medical & $\textbf{0.952 (0.894, 0.979)}$ & $\textbf{0.953 (0.906, 0.979)}$ & $\textbf{0.963 (0.939, 0.981)}$ & $\textbf{0.951 (0.899, 0.978)}$ & $\textbf{0.918 (0.857, 0.973)}$  \\ 
        ATB & 0.944 (0.881, 0.973) & 0.943 (0.892, 0.973) & 0.954 (0.926, 0.977) & 0.943 (0.889, 0.973) & 0.905 (0.835, 0.965)  \\ 
        Neurospin & 0.942 (0.874, 0.973) & 0.940 (0.888, 0.972) & 0.953 (0.924, 0.975) & 0.942 (0.887, 0.972) & 0.898 (0.824, 0.957)  \\ 
        \midrule \midrule
        \multicolumn{6}{@{}l}{\textbf{4X Transfer Track}}\\
        AIRS Medical & $\textbf{0.940 (0.802, 0.989)}$ & 0.902 (0.795, 0.983) & $\textbf{0.960 (0.942, 0.969)}$ & $\textbf{0.975 (0.934, 0.992)}$ & $\textbf{0.910 (0.728, 0.979)}$  \\ 
        MRRecon & 0.930 (0.846, 0.981) & $\textbf{0.946 (0.910, 0.975)}$ & 0.956 (0.935, 0.969) & 0.950 (0.889, 0.984) & 0.897 (0.671, 0.956)  \\ 
        ResoNNance & 0.913 (0.791, 0.977) & 0.936 (0.892, 0.967) & 0.939 (0.920, 0.956) & 0.957 (0.911, 0.980) & 0.855 (0.695, 0.951)  \\ 
        \bottomrule
    \end{tabular}
    \label{tab:resultsbycontrast}
\end{table*}
Table \ref{tab:resultsbycontrast} summarizes results by contrast for the finalists in each competition track with means and 95\% confidence intervals based on 2.5\% and 97.5\% quantiles.
The strongest SSIM scores were usually recorded on T1 post-contrast images (T1POST), while the weakest scores were typically on FLAIR images.
The same participant recorded the top average SSIM score for every contrast in every track except the Transfer track for T1 contrast.
In this case, two other participants posted higher SSIM scores.

\label{subsec:ssim_mask}
One team, HungryGrads, submitted to all tracks and received a very low SSIM score between 0.4 and 0.5.
This team set the background air to nearly 0s, which led to a clinically irrelevant SSIM loss of approximately 0.3 for their submissions.
The HungryGrads submission prompted our team to perform a post-hoc analysis where we masked both the submission and the reference RSS ground truth before calculating SSIM, with results plotted in Figure S1 in the supplementary material.
Applying this mask markedly improved the SSIM scores of HungryGrads, although it would not have made this team a finalist.
Applying the mask would have enabled ATB to enter the finalist round for the Transfer track.
Our custom mask would not have changed finalist rankings otherwise.

\subsection{Radiologist Evaluation Results}
\begin{figure}[htb]
    \centering
    \includegraphics[width=.45\textwidth]{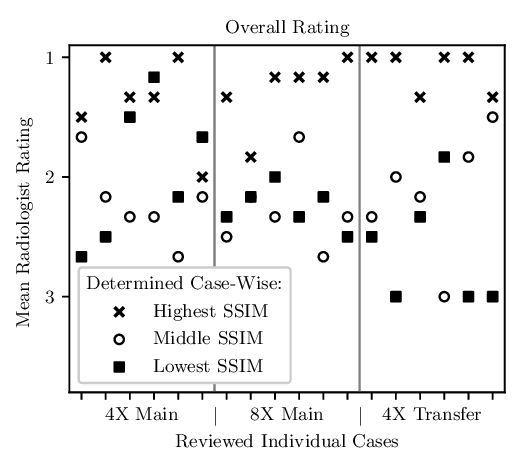}
    \caption{Scatter plot of mean radiologist rank across cases. The horizontal axis has a separate tick for each case evaluated by the radiologist cohort. The scatter plot markers indicate whether that method was from the team with the highest, middle, or lowest SSIM scores. We generally observed radiologists awarding the best ranks to models with the best SSIM score.}
    \label{fig:rad_ranking_case_wise}
\end{figure}

Radiologist rankings based on quality of pathology depiction were concordant with SSIM scores for the top submissions as shown in Figure \ref{fig:rad_ranking_case_wise}.
The second and third place performers for both 4X and 8X tracks were flipped between the quantitative ranking based on SSIM and the qualitative ranking based on radiologists.
The SSIM difference between these two constructions methods was relatively small, out to the third decimal place.
In the Transfer track, radiologist rankings matched ranking based on SSIM.

\begin{table}[htb]
    \centering
    \caption{Summary of Quality Ranks and Likert Scores}
    \begin{tabular}{lcccc@{}}
        \toprule
        Team         & Rank & Artifacts & Sharpness & CNR \\
        \midrule \midrule
        \multicolumn{5}{@{}l}{\textbf{4X Track}}\\
        AIRS & \textbf{1.36 $\boldsymbol{\pm}$ 0.64} & \textbf{1.53 $\boldsymbol{\pm}$ 0.70} & \textbf{1.53 $\boldsymbol{\pm}$ 0.51} & \textbf{1.53 $\boldsymbol{\pm}$ 0.51} \\
        Nspin    & 1.94 $\pm$ 0.86 & 1.81 $\pm$ 1.01 & 1.72 $\pm$ 0.66 & 1.75 $\pm$ 0.84 \\
        ATB          & 2.22 $\pm$ 0.87 & 1.75 $\pm$ 0.97 & 1.97 $\pm$ 0.65 & 1.86 $\pm$ 0.80 \\
        \midrule \midrule
        \multicolumn{5}{@{}l}{\textbf{8X Track}}\\
        AIRS & \textbf{1.28 $\boldsymbol{\pm}$ 0.64} & \textbf{1.67 $\boldsymbol{\pm}$ 0.68} & \textbf{1.89 $\boldsymbol{\pm}$ 0.75} & \textbf{1.94 $\boldsymbol{\pm}$ 0.75} \\
        Nspin    & 2.25 $\pm$ 0.77 & 1.86 $\pm$ 0.83 & 2.72 $\pm$ 0.81 & 2.28 $\pm$ 0.81 \\
        ATB          & 2.28 $\pm$ 0.70 & 1.92 $\pm$ 0.94 & 2.56 $\pm$ 0.77 & 2.42 $\pm$ 0.84 \\
        \midrule \midrule
        \multicolumn{5}{@{}l}{\textbf{4X Transfer Track}}\\
        AIRS & \textbf{1.11 $\boldsymbol{\pm}$ 0.32} & \textbf{1.42 $\boldsymbol{\pm}$ 0.50} & \textbf{1.83 $\boldsymbol{\pm}$ 0.65} & \textbf{1.81 $\boldsymbol{\pm}$ 0.62} \\
        MRR      & 1.97 $\pm$ 0.56 & 1.61 $\pm$ 0.55 & 2.41 $\pm$ 0.69 & 2.22 $\pm$ 0.64 \\
        Res.   & 2.78 $\pm$ 0.54 & 3.08 $\pm$ 0.84 & 2.86 $\pm$ 0.76 & 3.06 $\pm$ 0.86 \\
        \bottomrule
    \end{tabular}
    \label{tab:radresults}
\end{table}
A summary of the ranks and Likert scores with with means and standard deviations is shown in Table \ref{tab:radresults} (AIRS = AIRS Medical, Nspin = Neurospin, MRR = MRRecon, and Res. = ResoNNance).
We applied standard deviations for Table \ref{tab:radresults} (instead of quantiles as used in Table \ref{tab:resultsbycontrast}) to show more information on the variability.
Across all metrics AIRS Medical separated itself from the other submissions with the highest SSIM and best image quality.
Aside from this single team, differentiation among the other teams was not strong.
Of note, both the Neurospin and ATB teams had nearly identical average SSIM scores for the quantitative evaluation, with ATB presenting a slightly higher score (0.960 vs. 0.959 in 4X, 0.944 vs. 0.942 in 8X).
In the radiologist evaluation phase, these ranks flipped, with Neurospin receiving slightly higher ranks (1.94 vs. 2.22 in 4X, 2.25 vs. 2.28 in 8X).

A case-wise breakdown of the ranks for all 3 finalists and all rated cases is shown in Figure \ref{fig:rad_ranking_case_wise}.
For second and third-place metrics as rated by SSIM, radiologist assessment was discordant between the two methods.
However, in 16 out of 18 cases the highest SSIM score within the finalists' batches also received the highest radiologists' rating. A similar relation - not shown here - was found for the other used metrics such as normalized mean-squared error (NMSE) and peak signal-to-noise ratio (PSNR).

Radiologist agreement according to Kendall's coefficient of concordance generally improved as SSIM scores diverged.
We calculated the concordance using radiologist rankings of teams for quality of depiction of pathology vs. the ground truth.
For each case in the radiologist evaluation phase, we evaluated Kendall's coefficient of concordance with tie correction, and then aggregated over all cases by averaging.
This resulted in values of 0.457 for the 4X track, 0.386 for the 8X track, and 0.781 for the 4X Transfer track (where 0 indicates complete disagreement and 1 indicates complete agreement).
In the 4X and 8X tracks, discordance was primarily driven by two submissions (Neurospin and ATB) that were very close in SSIM score.
For the Transfer track, separation among the teams was more clear, and we observed corresponding increases in concordance.

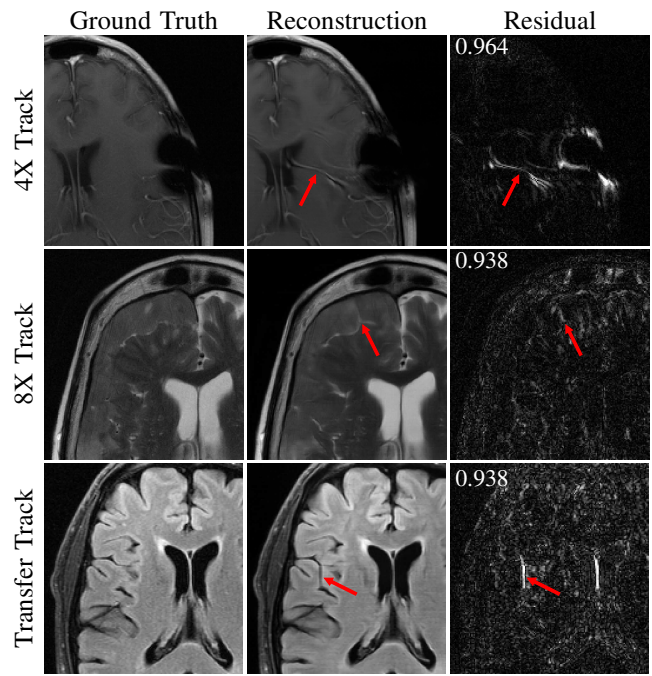
\begin{figure}[htb]
    \centering
    \begin{tikzpicture}
        \draw (0, 1.6cm) node[inner sep=0] {\raisebox{0in}{\rotatebox[origin=t]{90}{\mbox{}}}};
        \draw (1.6cm, 1.6cm) node[inner sep=0] (gtlabel) {Ground Truth};
        \draw (4.3cm, 1.6cm) node[inner sep=0] (gtlabel) {Reconstruction};
        \draw (7cm, 1.6cm) node[inner sep=0] (gtlabel) {Residual};
        \draw (0, 0cm) node[inner sep=0] {\raisebox{0in}{\rotatebox[origin=t]{90}{4X Track}}};
        \draw (1.6cm, 0cm) node[inner sep=0] {\includegraphics[height=2.8cm]{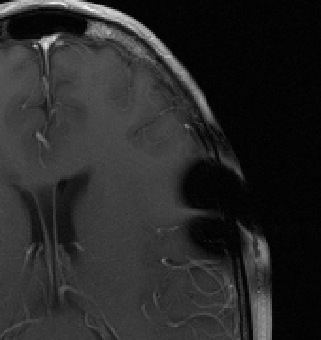}};
        \draw (4.3cm, 0cm) node[inner sep=0] {\includegraphics[height=2.8cm]{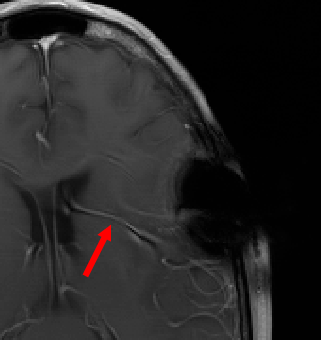}};
        \draw (7cm, 0cm) node[inner sep=0] {\includegraphics[height=2.8cm]{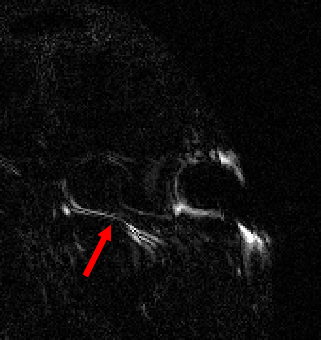}};
        \draw (6.1cm, 1.25cm) node[inner sep=0] {\small \color{white}0.964};
        \draw (0, -2.85cm) node[inner sep=0] {\raisebox{0in}{\rotatebox[origin=t]{90}{8X Track}}};
        \draw (1.6cm, -2.85cm) node[inner sep=0] {\includegraphics[height=2.8cm]{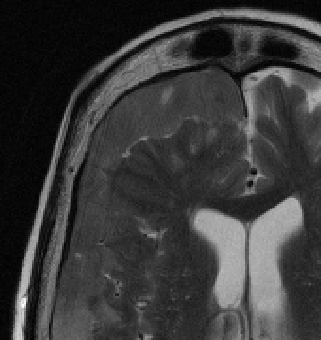}};
        \draw (4.3cm, -2.85cm) node[inner sep=0] {\includegraphics[height=2.8cm]{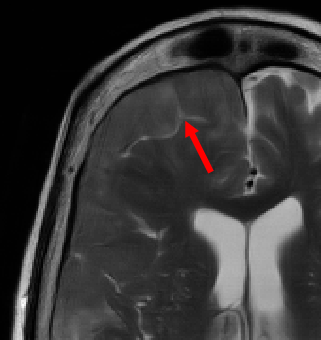}};
        \draw (7cm, -2.85cm) node[inner sep=0] {\includegraphics[height=2.8cm]{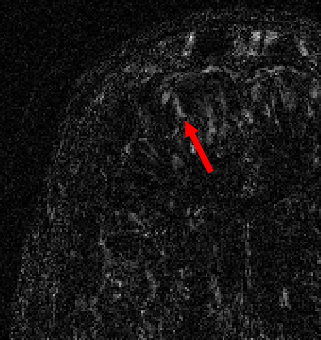}};
        \draw (6.1cm, -1.6cm) node[inner sep=0] {\small \color{white}0.938};
        \draw (0, -5.7cm) node[inner sep=0] {\raisebox{0in}{\rotatebox[origin=t]{90}{Transfer Track}}};
        \draw (1.6cm, -5.7cm) node[inner sep=0] {\includegraphics[height=2.8cm]{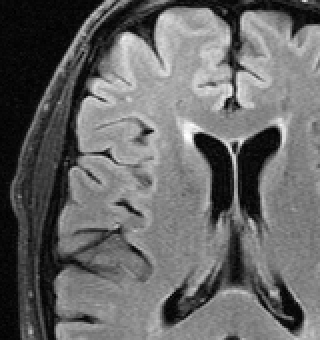}};
        \draw (4.3cm, -5.7cm) node[inner sep=0] {\includegraphics[height=2.8cm]{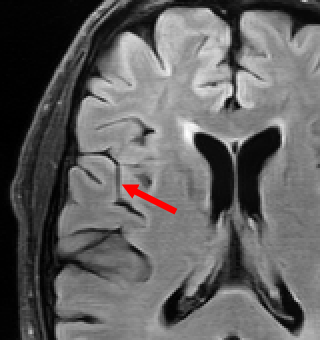}};
        \draw (7cm, -5.7cm) node[inner sep=0] {\includegraphics[height=2.8cm]{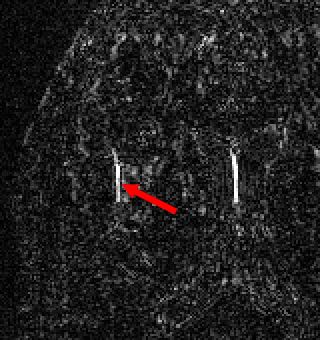}};
        \draw (6.1cm, -4.5cm) node[inner sep=0] {\small \color{white}0.938};
    \end{tikzpicture}
    \caption{Examples of reconstruction hallucinations among challenge submissions with SSIM scores over residual plots (residuals magnified by 5). (\textit{top}) A 4X submission from Neurospin generated a false vessel, possibly related to susceptibilities introduced by surgical staples. (\textit{middle}) An 8X submission from ATB introduced a linear bright signal mimicking a cleft of cerebrospinal fluid, as well as blurring of the boundaries of the extra-axial mass. (\textit{bottom}) A submission from ResoNNance introduced a false sulcus or prominent vessel.}
    \label{fig:hallucination}
\end{figure}

Radiologists did take note of hallucinatory effects introduced by the submission models.
Figure \ref{fig:hallucination} shows hallucination examples from all three tracks.
In some cases methods created artifact-mimics.
In other examples, models morphed an abnormality into a more normal brain structure, such as a sulcus or vessel.
Finally, we observed at least one example combining these two where an artifact was created at some intermediate layer of a model and then processed by the remaining portions of the network into a normal structure mimic.

\section{Discussion}
\subsection{Submission Overview}
In the 2019 challenge all three tracks were very closely contested, with little separation between teams either in the quantitative or the radiologist evaluation phases.
We observed this pattern to be reversed in the 2020 challenge, with one team assertively scoring the best in all evaluation phases.
For some images in the 4X track, multiple radiologists said that they did not observe major differentiating aspects affecting the depiction of pathology in the submissions.
However, when averaging the radiologists' rankings, radiologists preferred the method that had the highest-scoring on SSIM from AIRS Medical.
We further observed that the AIRS model scored highest on Likert-type ratings of artifacts, sharpness, and CNR.
This model also provided improvement over the baseline~\cite{sriram2020end}, which had previously been demonstrated for clinical interchangeability at 4X for knee imaging~\cite{recht2020using}.
Outside of the AIRS model, in the 4X and 8X tracks the second and third-place models scored very close together in both the quantitative and the qualitative evaluation phase.
In some cases the SSIM scores for these two models were identical out to three decimal places.

We observed decreases in performance in the Transfer track.
Many participants struggled to adapt their models to the GE data with its lack of disk-written frequency oversampling.
Although technically the GE scanner did not operate in any majorly different way than Philips and Siemens scanners (all use frequency oversampling), this simple aspect rendered many models useless in this track without modification.
Another factor was a divergence in FLAIR methodology: our Philips and Siemens data used T2 FLAIR images, whereas the GE data had T1 FLAIR images.
Modifications for correcting these effects seem not to be straightforward.
We note that as designed the Transfer track primarily evaluated one type of transfer: generalization across vendors. This was the most commonly-cited type of transfer in feedback from the 2019 challenge, but future challenges may investigate other types of transfer.

In terms of radiologist evaluations, despite the drawbacks of SSIM and RSS ground truths, we observed a correlation between radiologist scores and SSIM scores for large SSIM separations.
Multiple radiologists found images at 4X to be similar in terms of depiction of the pathology, although artifacts tended to be more problematic in T1POST images.
When it came to the 8X and Transfer tracks, radiologist sentiment became more negative.
Multiple radiologists in both  of these tracks offered feedback that none of the submitted images would be acceptable, indicating that these two tasks may remain open problems going into 2021.

We note that the results of this paper were observed within the regime of retrospective undersampling. Retrospective undersampling does not consider potential differences in signal relaxation along the echo trains. Even though echo trains can be designed for most sequences in such a manner that undersampling does not lead to a change in overall relaxation weighting along the phase dimension, we have not shown equivalency in our work. We would recommend that researchers confirm the results of the methods in this paper with prospective sampling prior to clinical use.

The results of the challenge suggests a few conclusions on approaches.
The first: cascaded models with a data fidelity term and CNN regularization continue to dominate the submission field, as occurred in the previous challenge~\cite{Knoll2019}.
Second: AIRS, the team that won all three tracks, had the largest model.
However, large models were not always better, with ATB having a model with similar performance to Neurospin despite having 87\% less parameters.
Lastly, we note the AIRS model used a normalization routine to get the data into a consistent format for all coil configurations, as well as being the only team to use a GRAPPA~\cite{griswold2002generalized} reconstruction as the initialization.

\subsection{Quantitative Evaluation Process}
Discussions around the quantitative evaluation process primarily concerned the presence of background noise during both the planning and execution stage of the challenge.
The influence of background noise on SSIM scores is substantial.
One participant in the 2019 challenge had a dedicated style transfer model in order to add this noise back into the reconstructed images~\cite{hammernik2019sigma}.
Despite the drawbacks to SSIM, we were unable to agree on an alternative for the 2020 challenge.

In the 2020 challenge, the HungryGrads team submitted images with backgrounds of nearly zeroes, which penalized their scores.
Prompted by this submission we investigated the effect a masked metric might have had on their scores in Section \ref{subsec:ssim_mask}.
We opted for a masking algorithm that removes most background pixels and altered the algorithm parameters for low-SNR edge cases where it did not perform well.
Due to the relatively small size of the challenge data set visual inspection of the validity of the masks was feasible.
The ranking of the challenge did not change dramatically due to masking, but masking made metrics less prone to a specific reconstruction method's impact on the background. 

Another intriguing alternative would be to use alternative reconstruction techniques such as adaptive combine that implicitly suppress background noise~\cite{walsh2000adaptive}.
We considered using adaptive combine reconstructions in our evaluation, but concluded that the results were not particularly meaningful as most models had been trained with RSS backgrounds.
Another alternative to adaptive combine would be to use other metrics that are more aware of the noise properties, such as Stein’s Unbiased Risk Estimate (SURE).

One area lacking in our quantitative analysis was hallucination detection.
This is an area of great interest to the community, but as of the end of our challenge we were unaware of automated, quantitative methods for detecting lesions or characterizing stability (although some methods can demonstrate instability qualitatively~\cite{antun2020instabilities}).
Detection of automated stability/hallucination analysis remains a topic of great interest for future challenges.

\subsection{Qualitative Radiologist Evaluation}
For the 2020 challenge, we altered the radiologist questionnaire to focus their ranking on the depiction of pathologies rather than general image quality.
Some radiologists found the focus helpful, commenting specifically that the images at 8X and in the 4X Transfer track might not be acceptable for clinical use.
As this task aligns more closely with the normal clinical workflow, we would encourage future competition organizers to use this approach for their radiologist evaluation procedures.

In the 4X track, there were specific cases where the radiologist rankings were concordant and others where the rankings were discordant.
Discordant cases tended, upon review, to show that the main abnormalities were similarly well depicted across the top 3 reconstructions, though there were oftentimes concordant estimations of differences between reconstructions in terms of artifacts, sharpness and CNR.

Radiologist sentiment was affected by hallucinations such as those in Figure \ref{fig:hallucination}.
Such hallucinatory features are not acceptable and especially problematic if they mimic normal structures that are either not present or actually abnormal.
These images had high SSIM scores, indicating that even though these images are considered well-optimized according to this metric, they are not optimized regarding hallucination features.
Neural network models can be unstable as demonstrated via adversarial perturbation studies~\cite{antun2020instabilities}.
Despite the lack of realism in some of these perturbations, our results indicate that hallucination and artifacts remain a real concern, particularly at higher accelerations.
This topic is in major need for further development.

We note that we did not perform intensity correction for either the participant submissions or the ground truth reconstructions. None of the radiologists commented on intensity inhomogeneity in either the initial testing phase or the final evaluation phase. However, intensity correction is routinely done by vendor scanners and could have affected the evaluation, as some pathologies manifest via varying tissue intensities.

\subsection{Feedback from Participants}
We asked participants for feedback regarding challenge organization and future organization.
Participants were generally enthusiastic about being able to participate in the challenge.
We received positive feedback on our communication via the fastMRI GitHub repository at \url{https://github.com/facebookresearch/fastMRI} and the forum associated with the web site at \url{https://fastmri.org}.
We also received positive feedback around the challenge's realism in focusing on multi-coil data, as well as the challenge's generalizability initiative in focusing on the Transfer track.

Still, participants felt the realism could be improved in other areas.
In particular, the sampling mask used for the challenge used pseudo-regular sampling in order to achieve exact 4X and 8X sampling rates.
This sampling pattern is not equivalent to the perfectly equidistant sampling pattern used on MRI systems, which give acceleration rates slightly less than the target rate due to the densely-sampled center.
As a result, challenge models are likely to require further fine-tuning training before application to clinical data.

Another point of feedback centered on the storage and compute resources necessary to participate in the challenge.
In the 2019 challenge, the storage aspect was mitigated by the inclusion of the single-coil track (which had a smaller download size).
The single-coil track attracted a lot of engagement, with 25 out of 33 groups submitting to it~\cite{knoll2020advancing}.
From the compute angle, the trend towards larger models requires costly hardware.
Training the baseline End-to-End Variational Network~\cite{sriram2020end} requires 32 GPUs, each with 32 GB of memory, for about 3.5 days.
This level of compute power is not available at many academic centers.
By comparison, multiple participants submitted models trained on only a single GPU.
This was also a topic of feedback from non-participants, with some telling us informally that they did not participate due to compute or storage requirements.
For the future, researchers felt it would be helpful for the barriers to entry were lower, particularly for academic groups that might have innovative methods but less compute or storage.

As always, the selection of best quantitative evaluation metrics to use is extremely difficult and there are potential drawbacks to many or all.
Participants did provide feedback concerning the use of SSIM and the use of RSS for ground truth images.
Although groups acknowledged efforts to seek superior metrics, they felt that settling for this particular metric was disappointing.
Some participants felt there was a tradeoff between optimizing for SSIM (which promotes smoothing) vs. radiologist interpretation.
Most vendors have variations in their post-processing pipelines for precisely this reason.
Some vendor post-processing methods even allow for radiologists to adjust the strength of the regularization.
We did not allow secondary submissions from participants that might enhance the images for human perception, such as those based on noise dithering or inspired by stochastic resonance~\cite{moss2004stochastic,recht2020using}.
Allowing secondary submissions for radiologist interpretation may be beneficial for future challenges, provided ground truth images are also included to allow radiologists to watch for hallucination.
In compiling the results for this challenge we have attempted to investigate some other options that would at least mitigate the effects of background noise and feel that this is an important topic for further investigation.
Consensus around evaluations--for ground truth calculations, metrics, and radiologist presentation--would substantially aid the organization of future challenges.

\section{Conclusion}
The 2020 fastMRI reconstruction challenge featured two core modifications from its 2019 predecessor: 1) a new competition Transfer track to evaluate model generalization and 2) adjusting the radiologist evaluation to focus on pathology depiction.
In addition to these, we extended our competition to a new anatomy with much larger data sets for both training and competition evaluation.
The competition resulted in a new state-of-the-art model.
Our challenge confirmed areas in need of research, particularly those along the lines of evaluation metrics, error characterization, and AI-generated hallucinations.
Radiologist sentiment was mixed for images submitted to the 8X and the Transfer tracks; these may remain open research frontiers going into 2021.
We hope that researchers and future challenge organizers find the results of the 2020 fastMRI challenge helpful in their future endeavors.

\section*{Acknowledgment}
We would like to thank all research groups that participated in the challenge.
We thank the scientific advisors that gave feedback during challenge organization.
We thank Girish M. Fatterpekar, Nael Kambiz, Amy L. Kotsenas, Christie Mary Lincoln, Joseph Mettenburg, and Max Wintermark for judging the submissions and the contribution of their clinical expertise and time.
We would also like to thank Tullie Murrell, Mark Tygert, and C. Lawrence Zitnick for many valuable contributions to the launch and execution of this challenge.
We give special thanks to Philips North America and their clinical partner sites for the data they contributed for use as part of the Transfer track.
The NYU authors thank NIH grants P41EB017183, R01EB024532, and R21EB027241 for financial support.
The ResoNNance team's submission was conducted under the STAIRS project under the TKI-PPP program.
The ResoNNance collaboration project is co-funded by the PPP Allowance made available by Health Holland, Top Sector Life Sciences \& Health, to stimulate public-private partnerships.

\ifCLASSOPTIONcaptionsoff
  \newpage
\fi

\bibliographystyle{IEEEtran}
\bibliography{references}

\begin{thebibliography}{10}
\providecommand{\url}[1]{#1}
\csname url@samestyle\endcsname
\providecommand{\newblock}{\relax}
\providecommand{\bibinfo}[2]{#2}
\providecommand{\BIBentrySTDinterwordspacing}{\spaceskip=0pt\relax}
\providecommand{\BIBentryALTinterwordstretchfactor}{4}
\providecommand{\BIBentryALTinterwordspacing}{\spaceskip=\fontdimen2\font plus
\BIBentryALTinterwordstretchfactor\fontdimen3\font minus
  \fontdimen4\font\relax}
\providecommand{\BIBforeignlanguage}[2]{{%
\expandafter\ifx\csname l@#1\endcsname\relax
\typeout{** WARNING: IEEEtran.bst: No hyphenation pattern has been}%
\typeout{** loaded for the language `#1'. Using the pattern for}%
\typeout{** the default language instead.}%
\else
\language=\csname l@#1\endcsname
\fi
#2}}
\providecommand{\BIBdecl}{\relax}
\BIBdecl

\bibitem{chetlur2014cudnn}
S.~Chetlur, C.~Woolley, P.~Vandermersch, J.~Cohen, J.~Tran, B.~Catanzaro, and
  E.~Shelhamer, ``{cuDNN}: Efficient primitives for deep learning,''
  \emph{arXiv preprint arXiv:1410.0759}, 2014.

\bibitem{abadi2016tensorflow}
M.~Abadi, P.~Barham, J.~Chen, Z.~Chen, A.~Davis, J.~Dean, M.~Devin,
  S.~Ghemawat, G.~Irving \emph{et~al.}, ``{TensorFlow}: A system for
  large-scale machine learning,'' in \emph{Proc. USENIX Conf. Operating Sys.
  Design and Impl.}, 2016, pp. 265--283.

\bibitem{paszke2019pytorch}
A.~Paszke, S.~Gross, F.~Massa, A.~Lerer, J.~Bradbury, G.~Chanan, T.~Killeen,
  Z.~Lin, N.~Gimelshein \emph{et~al.}, ``{PyTorch}: An imperative style,
  high-performance deep learning library,'' in \emph{Adv. Neural Inf. Process.
  Sys.}, 2019, pp. 8026--8037.

\bibitem{sun2016deep}
Y.~Yang, J.~Sun, H.~Li, and Z.~Xu, ``Deep {ADMM-Net} for compressive sensing
  {MRI},'' in \emph{Adv. Neural Inf. Process. Sys.}, 2016, pp. 10--18.

\bibitem{hammernik2018learning}
K.~Hammernik, T.~Klatzer, E.~Kobler, M.~P. Recht, D.~K. Sodickson, T.~Pock, and
  F.~Knoll, ``Learning a variational network for reconstruction of accelerated
  {MRI} data,'' \emph{Magn. Res. Med.}, vol.~79, no.~6, pp. 3055--3071, 2018.

\bibitem{schlemper2017deep}
J.~Schlemper, J.~Caballero, J.~V. Hajnal, A.~N. Price, and D.~Rueckert, ``A
  deep cascade of convolutional neural networks for dynamic {MR} image
  reconstruction,'' \emph{IEEE Trans. Med. Imag.}, vol.~37, no.~2, pp.
  491--503, 2017.

\bibitem{yang2017dagan}
G.~Yang, S.~Yu, H.~Dong, G.~Slabaugh, P.~L. Dragotti, X.~Ye, F.~Liu,
  S.~Arridge, J.~Keegan \emph{et~al.}, ``{DAGAN}: Deep de-aliasing generative
  adversarial networks for fast compressed sensing {MRI} reconstruction,''
  \emph{IEEE Trans. Med. Imag.}, vol.~37, no.~6, pp. 1310--1321, 2017.

\bibitem{eo2018kiki}
T.~Eo, Y.~Jun, T.~Kim, J.~Jang, H.-J. Lee, and D.~Hwang, ``{KIKI-net}:
  Cross-domain convolutional neural networks for reconstructing undersampled
  magnetic resonance images,'' \emph{Magn. Res. Med.}, vol.~80, no.~5, pp.
  2188--2201, 2018.

\bibitem{aggarwal2018modl}
H.~K. Aggarwal, M.~P. Mani, and M.~Jacob, ``{MoDL}: Model-based deep learning
  architecture for inverse problems,'' \emph{IEEE Trans. Med. Imag.}, vol.~38,
  no.~2, pp. 394--405, 2018.

\bibitem{zhu2018image}
B.~Zhu, J.~Z. Liu, S.~F. Cauley, B.~R. Rosen, and M.~S. Rosen, ``Image
  reconstruction by domain-transform manifold learning,'' \emph{Nature}, vol.
  555, no. 7697, pp. 487--492, 2018.

\bibitem{Knoll2019}
F.~Knoll, K.~Hammernik, E.~Kobler, T.~Pock, M.~P. Recht, and D.~K. Sodickson,
  ``{Assessment of the generalization of learned image reconstruction and the
  potential for transfer learning},'' \emph{Magn. Res. Med.}, vol.~81, no.~1,
  pp. 116--128, 2019.

\bibitem{Knoll2020IEEE}
F.~Knoll, K.~Hammernik, C.~Zhang, S.~Moeller, T.~Pock, D.~K. Sodickson, and
  M.~Ak{\c{c}}akaya, ``Deep-learning methods for parallel magnetic resonance
  imaging reconstruction: A survey of the current approaches, trends, and
  issues,'' \emph{IEEE Signal Process. Mag.}, vol.~37, no.~1, pp. 128--140,
  2020.

\bibitem{yaman2020self}
B.~Yaman, S.~A.~H. Hosseini, S.~Moeller, J.~Ellermann, K.~U{\u{g}}urbil, and
  M.~Ak{\c{c}}akaya, ``Self-supervised learning of physics-guided
  reconstruction neural networks without fully sampled reference data,''
  \emph{Magn. Res. Med.}, vol.~84, no.~6, pp. 3172--3191, 2020.

\bibitem{sriram2020end}
A.~Sriram, J.~Zbontar, T.~Murrell, A.~Defazio, C.~L. Zitnick, N.~Yakubova,
  F.~Knoll, and P.~Johnson, ``End-to-end variational networks for accelerated
  {MRI} reconstruction,'' in \emph{Intl. Conf. Med. Imag. Comput.
  Comput.-Assisted Intervention}, 2020, pp. 64--73.

\bibitem{fukushima1980neocognitron}
K.~Fukushima, ``Neocognitron: A self-organizing neural network model for a
  mechanism of pattern recognition unaffected by shift in position,''
  \emph{Biol. Cybern.}, vol.~36, no.~4, pp. 193--202, 1980.

\bibitem{lecun1998gradient}
Y.~LeCun, L.~Bottou, Y.~Bengio, and P.~Haffner, ``Gradient-based learning
  applied to document recognition,'' \emph{Proc. IEEE}, vol.~86, no.~11, pp.
  2278--2324, 1998.

\bibitem{raina2009large}
R.~Raina, A.~Madhavan, and A.~Y. Ng, ``Large-scale deep unsupervised learning
  using graphics processors,'' in \emph{Proc. Intl. Conf. Mach. Learning},
  2009, pp. 873--880.

\bibitem{cirecsan2010deep}
D.~C. Cire{\c{s}}an, U.~Meier, L.~M. Gambardella, and J.~Schmidhuber, ``Deep,
  big, simple neural nets for handwritten digit recognition,'' \emph{Neural
  Comput.}, vol.~22, no.~12, pp. 3207--3220, 2010.

\bibitem{deng2009imagenet}
J.~Deng, W.~Dong, R.~Socher, L.-J. Li, K.~Li, and L.~Fei-Fei, ``{ImageNet}: A
  large-scale hierarchical image database,'' in \emph{Proc. IEEE Conf. Comput.
  Vis. Pattern Recogn.}, 2009, pp. 248--255.

\bibitem{krizhevsky2012imagenet}
A.~Krizhevsky, I.~Sutskever, and G.~E. Hinton, ``{ImageNet} classification with
  deep convolutional neural networks,'' in \emph{Adv. Neural Inf. Process.
  Sys.}, 2012, pp. 1097--1105.

\bibitem{simonyan2014very}
K.~Simonyan and A.~Zisserman, ``Very deep convolutional networks for
  large-scale image recognition,'' in \emph{Intl. Conf. Learning Repr.}, 2015.

\bibitem{he2016deep}
K.~He, X.~Zhang, S.~Ren, and J.~Sun, ``Deep residual learning for image
  recognition,'' in \emph{Proc. IEEE Conf. Comput. Vis. Pattern Recogn.}, 2016,
  pp. 770--778.

\bibitem{szegedy2016rethinking}
C.~Szegedy, V.~Vanhoucke, S.~Ioffe, J.~Shlens, and Z.~Wojna, ``Rethinking the
  {Inception} architecture for computer vision,'' in \emph{Proc. IEEE Conf.
  Comput. Vis. Pattern Recogn.}, 2016, pp. 2818--2826.

\bibitem{huang2017densely}
G.~Huang, Z.~Liu, L.~Van Der~Maaten, and K.~Q. Weinberger, ``Densely connected
  convolutional networks,'' in \emph{Proc. IEEE Conf. Comput. Vis. Pattern
  Recogn.}, 2017, pp. 4700--4708.

\bibitem{zbontar2018fastmri}
J.~Zbontar, F.~Knoll, A.~Sriram, T.~Murrell, Z.~Huang, M.~J. Muckley,
  A.~Defazio, R.~Stern, P.~Johnson \emph{et~al.}, ``{fastMRI}: An open dataset
  and benchmarks for accelerated {MRI},'' \emph{arXiv preprint
  arXiv:1811.08839}, 2018.

\bibitem{knoll2020fastmri}
F.~Knoll, J.~Zbontar, A.~Sriram, M.~J. Muckley, M.~Bruno, A.~Defazio,
  M.~Parente, K.~J. Geras, J.~Katsnelson \emph{et~al.}, ``{fastMRI}: A publicly
  available raw k-space and {DICOM} dataset of knee images for accelerated {MR}
  image reconstruction using machine learning,'' \emph{Radiol.: Artificial
  Intell.}, vol.~2, no.~1, p. e190007, 2020.

\bibitem{knoll2020advancing}
F.~Knoll, T.~Murrell, A.~Sriram, N.~Yakubova, J.~Zbontar, M.~Rabbat,
  A.~Defazio, M.~J. Muckley, D.~K. Sodickson \emph{et~al.}, ``Advancing machine
  learning for {MR} image reconstruction with an open competition: Overview of
  the 2019 {fastMRI} challenge,'' \emph{Magn. Res. Med.}, vol.~84, no.~6, pp.
  3054--3070, 2020.

\bibitem{grissom2017advancing}
W.~A. Grissom, K.~Setsompop, S.~A. Hurley, J.~Tsao, J.~V. Velikina, and A.~A.
  Samsonov, ``Advancing {RF} pulse design using an open-competition format:
  Report from the 2015 {ISMRM} challenge,'' \emph{Magn. Res. Med.}, vol.~78,
  no.~4, pp. 1352--1361, 2017.

\bibitem{schilling2019challenges}
K.~G. Schilling, A.~Daducci, K.~Maier-Hein, C.~Poupon, J.-C. Houde, V.~Nath,
  A.~W. Anderson, B.~A. Landman, and M.~Descoteaux, ``Challenges in diffusion
  {MRI} tractography--lessons learned from international benchmark
  competitions,'' \emph{Magn. Res. Imag.}, vol.~57, pp. 194--209, 2019.

\bibitem{maier2020cg}
O.~Maier, S.~H. Baete, A.~Fyrdahl, K.~Hammernik, S.~Harrevelt, L.~Kasper,
  A.~Karakuzu, M.~Loecher, F.~Patzig \emph{et~al.}, ``{CG‐SENSE} revisited:
  Results from the first {ISMRM} reproducibility challenge,'' \emph{Magn. Res.
  Med.}, 2020.

\bibitem{beauferris2020multichannel}
Y.~Beauferris, J.~Teuwen, D.~Karkalousos, N.~Moriakov, M.~Caan, L.~Rodrigues,
  A.~Lopes, H.~Pedrini, L.~Rittner \emph{et~al.}, ``Multi-channel {MR}
  reconstruction ({MC-MRRec}) challenge -- comparing accelerated {MR}
  reconstruction models and assessing their generalizability to datasets
  collected with different coils,'' \emph{arXiv preprint arXiv:2011.07952},
  2020.

\bibitem{whittaker1915xviii}
E.~T. Whittaker, ``{XVIII.}—{On} the functions which are represented by the
  expansions of the interpolation-theory,'' \emph{Proc. Royal Soc. Edinburgh},
  vol.~35, pp. 181--194, 1915.

\bibitem{nyquist1928certain}
H.~Nyquist, ``Certain topics in telegraph transmission theory,'' \emph{Trans.
  Am. Inst. Electr. Eng.}, vol.~47, no.~2, pp. 617--644, 1928.

\bibitem{kotel1933carrying}
V.~A. Kotelnikov, ``On the carrying capacity of the `ether' and wire in
  telecommunications,'' in \emph{Mat. First All-Union Conf. on Questions of
  Comm. (Russian)}, 1933.

\bibitem{shannon1949communication}
C.~E. Shannon, ``Communication in the presence of noise,'' \emph{Proc. IRE},
  vol.~37, no.~1, pp. 10--21, 1949.

\bibitem{sodickson1997simultaneous}
D.~K. Sodickson and W.~J. Manning, ``Simultaneous acquisition of spatial
  harmonics ({SMASH}): Fast imaging with radiofrequency coil arrays,''
  \emph{Magn. Res. Med.}, vol.~38, no.~4, pp. 591--603, 1997.

\bibitem{pruessmann1999sense}
K.~P. Pruessmann, M.~Weiger, M.~B. Scheidegger, and P.~Boesiger, ``{SENSE}:
  Sensitivity encoding for fast {MRI},'' \emph{Magn. Res. Med.}, vol.~42,
  no.~5, pp. 952--962, 1999.

\bibitem{griswold2002generalized}
M.~A. Griswold, P.~M. Jakob, R.~M. Heidemann, M.~Nittka, V.~Jellus, J.~Wang,
  B.~Kiefer, and A.~Haase, ``Generalized autocalibrating partially parallel
  acquisitions ({GRAPPA}),'' \emph{Magn. Res. Med.}, vol.~47, no.~6, pp.
  1202--1210, 2002.

\bibitem{lustig2008compressed}
M.~Lustig, D.~L. Donoho, J.~M. Santos, and J.~M. Pauly, ``Compressed sensing
  {MRI},'' \emph{IEEE Sig. Process. Mag.}, vol.~25, no.~2, pp. 72--82, 2008.

\bibitem{uecker2014espirit}
M.~Uecker, P.~Lai, M.~J. Murphy, P.~Virtue, M.~Elad, J.~M. Pauly, S.~S.
  Vasanawala, and M.~Lustig, ``{ESPIRiT}—an eigenvalue approach to
  autocalibrating parallel {MRI}: Where {SENSE} meets {GRAPPA},'' \emph{Magn.
  Res. Med.}, vol.~71, no.~3, pp. 990--1001, 2014.

\bibitem{ronneberger2015u}
O.~Ronneberger, P.~Fischer, and T.~Brox, ``{U-Net}: Convolutional networks for
  biomedical image segmentation,'' in \emph{Intl. Conf. Med. Imag. Comput.
  Comput.-Assisted Intervention}, 2015, pp. 234--241.

\bibitem{pezzotti2020adaptive}
N.~Pezzotti, S.~Yousefi, M.~S. Elmahdy, J.~V. Gemert, C.~Schulke, M.~Doneva,
  T.~Nielsen, S.~Kastryulin, B.~P.~F. Lelieveldt \emph{et~al.}, ``An adaptive
  intelligence algorithm for undersampled knee {MRI} reconstruction,''
  \emph{IEEE Access}, vol.~8, pp. 204\,825--204\,838, 2020.

\bibitem{wang2019pyramid}
P.~Wang, E.~Z. Chen, T.~Chen, V.~M. Patel, and S.~Sun, ``Pyramid convolutional
  {RNN} for {MRI} reconstruction,'' \emph{arXiv preprint arXiv:1912.00543},
  2019.

\bibitem{putzky2019rim}
P.~Putzky, D.~Karkalousos, J.~Teuwen, N.~Miriakov, B.~Bakker, M.~Caan, and
  M.~Welling, ``i-{RIM} applied to the {fastMRI} challenge,'' \emph{arXiv
  preprint arXiv:1910.08952}, 2019.

\bibitem{wang2004image}
Z.~Wang, A.~C. Bovik, H.~R. Sheikh, and E.~P. Simoncelli, ``Image quality
  assessment: from error visibility to structural similarity,'' \emph{IEEE
  Trans. Imag. Process.}, vol.~13, no.~4, pp. 600--612, 2004.

\bibitem{roemer1990nmr}
P.~B. Roemer, W.~A. Edelstein, C.~E. Hayes, S.~P. Souza, and O.~M. Mueller,
  ``The {NMR} phased array,'' \emph{Magn. Res. Med.}, vol.~16, no.~2, pp.
  192--225, 1990.

\bibitem{walsh2000adaptive}
D.~O. Walsh, A.~F. Gmitro, and M.~W. Marcellin, ``Adaptive reconstruction of
  phased array {MR} imagery,'' \emph{Magn. Res. Med.}, vol.~43, no.~5, pp.
  682--690, 2000.

\bibitem{recht2020using}
M.~P. Recht, J.~Zbontar, D.~K. Sodickson, F.~Knoll, N.~Yakubova, A.~Sriram,
  T.~Murrell, A.~Defazio, M.~Rabbat \emph{et~al.}, ``Using deep learning to
  accelerate knee {MRI} at {3T}: Results of an interchangeability study,''
  \emph{Am. J. Roentgenology}, vol. 215, no.~6, pp. 1421--1429, 2020.

\bibitem{van2014scikit}
S.~Van~der Walt, J.~L. Sch{\"o}nberger, J.~Nunez-Iglesias, F.~Boulogne, J.~D.
  Warner, N.~Yager, E.~Gouillart, and T.~Yu, ``scikit-image: image processing
  in {Python},'' \emph{PeerJ}, vol.~2, p. e453, 2014.

\bibitem{Kingma2015AdamAM}
D.~P. Kingma and J.~Ba, ``Adam: A method for stochastic optimization,'' in
  \emph{Intl. Conf. Learning Repr.}, 2015.

\bibitem{jun2021joint}
Y.~Jun, H.~Shin, and D.~Hwang, ``Joint deep model-based {MR} image and coil
  sensitivity reconstruction network ({Joint-ICNet}) for fast {MRI},'' in
  \emph{Proc. IEEE/CVF Conf. Comput. Vis. Pattern Recogn.}, June 2021, in
  press.

\bibitem{Yu_2019_CVPR_Workshops}
S.~Yu, B.~Park, and J.~Jeong, ``Deep iterative down-up {CNN} for image
  denoising,'' in \emph{Proc. IEEE/CVF Conf. Comput. Vis. Pattern Recogn.
  Workshops}, June 2019.

\bibitem{ramzi2020benchmarking}
Z.~Ramzi, P.~Ciuciu, and J.-L. Starck, ``Benchmarking {MRI} reconstruction
  neural networks on large public datasets,'' \emph{App. Sci.}, vol.~10, no.~5,
  p. 1816, 2020.

\bibitem{ramzi2021xpdnet}
Z.~Ramzi, J.-L. Starck, and P.~Ciuciu, ``{XPDNet} for {MRI} reconstruction: An
  application to the 2020 {fastMRI} challenge,'' in \emph{Proc. Intl. Soc.
  Magn. Res. Med.}, 2021, in press.

\bibitem{chambolle2011first}
A.~Chambolle and T.~Pock, ``A first-order primal-dual algorithm for convex
  problems with applications to imaging,'' \emph{J. of Mathematical Imag.
  Vis.}, vol.~40, no.~1, pp. 120--145, 2011.

\bibitem{adler2018learned}
J.~Adler and O.~{\"O}ktem, ``Learned primal-dual reconstruction,'' \emph{IEEE
  Trans. Med. Imag.}, vol.~37, no.~6, pp. 1322--1332, 2018.

\bibitem{liu2018multi}
P.~Liu, H.~Zhang, K.~Zhang, L.~Lin, and W.~Zuo, ``Multi-level {Wavelet-CNN} for
  image restoration,'' in \emph{Proc. IEEE Conf. Comput. Vis. Pattern Recogn.
  Workshops}, 2018, pp. 773--782.

\bibitem{Liu2020OnTV}
L.~Liu, H.~Jiang, P.~He, W.~Chen, X.~Liu, J.~Gao, and J.~Han, ``On the variance
  of the adaptive learning rate and beyond,'' in \emph{Intl. Conf. Learning
  Repr.}, 2020.

\bibitem{putzky2017recurrent}
P.~Putzky and M.~Welling, ``Recurrent inference machines for solving inverse
  problems,'' \emph{arXiv preprint arXiv:1706.04008}, 2017.

\bibitem{lonning2019recurrent}
K.~L{\o}nning, P.~Putzky, J.-J. Sonke, L.~Reneman, M.~W. Caan, and M.~Welling,
  ``Recurrent inference machines for reconstructing heterogeneous {MRI} data,''
  \emph{Med. Imag. Anal.}, vol.~53, pp. 64--78, 2019.

\bibitem{uecker2015berkeley}
M.~Uecker, F.~Ong, J.~I. Tamir, D.~Bahri, P.~Virtue, J.~Y. Cheng, T.~Zhang, and
  M.~Lustig, ``Berkeley advanced reconstruction toolbox,'' in \emph{Proc. Intl.
  Soc. Magn. Res. Med.}, 2015.

\bibitem{defazio2020mri}
A.~Defazio, T.~Murrell, and M.~P. Recht, ``{MRI} banding removal via
  adversarial training,'' in \emph{Adv. Neural Inf. Process. Sys.}, 2020, pp.
  7660--7670.

\bibitem{hammernik2019sigma}
K.~Hammernik, J.~Schlemper, C.~Qin, J.~Duan, R.~M. Summers, and D.~Rueckert,
  ``{$\Sigma$}-net: Systematic evaluation of iterative deep neural networks for
  fast parallel {MR} image reconstruction,'' \emph{arXiv preprint
  arXiv:1912.09278}, 2019.

\bibitem{antun2020instabilities}
V.~Antun, F.~Renna, C.~Poon, B.~Adcock, and A.~C. Hansen, ``On instabilities of
  deep learning in image reconstruction and the potential costs of {AI},''
  \emph{Proc. Natl. Acad. Sci.}, vol. 117, no.~48, pp. 30\,088--30\,095, 2020.

\bibitem{moss2004stochastic}
F.~Moss, L.~M. Ward, and W.~G. Sannita, ``Stochastic resonance and sensory
  information processing: A tutorial and review of application,'' \emph{Clin.
  Neurophys.}, vol. 115, no.~2, pp. 267--281, 2004.

\end{thebibliography}

\end{document}


\maketitle

Here we include supplementary material for the paper, ``Results of the 2020 fastMRI Challenge for Machine Learning MR Image Reconstruction.''

\section{Summary of Data Statistics}
Table \ref{tab:dataset_summary} shows a summary of the data used in the challenge. The Siemens data set was previously reported in an update of the fastMRI arXiv paper \cite{zbontar2018fastmri}. The GE data set and Philips data sets were collected specifically for the challenge.
\begin{table*}[htb]
    \small
    \centering
    \caption{Summary of Challenge Data}
    \begin{tabular}{lP{20mm}P{20mm}P{20mm}P{20mm}@{}}
        \toprule
        Parameter         & T1    & T1POST & T2   & FLAIR \\
        \midrule \midrule
        \multicolumn{5}{@{}l}{\textbf{Siemens Data Set}}\\
        Field of View (mm) & 220$\times$178 - 240$\times$240 & 220$\times$178 - 240$\times$240 & 220$\times$165 - 230$\times$230 & 200$\times$162 - 230$\times$230 \\ 
        Matrix Size & 320$\times$260 - 320$\times$330 & 320$\times$260 - 320$\times$320 & 320$\times$240 - 384$\times$384 & 256$\times$208 - 512$\times$512 \\ 
        Slice Thickness (mm) & 5 & 3-5 & 5 & 3-5 \\ 
        Number of Slices & 2-16 & 10-16 & 10-16 & 12-16 \\ 
        TR (ms) & 250-786 & 247-786 & 4000-15810 & 9000 \\ 
        TE (ms) & 2-9 & 2-11 & 102-115 & 76-126 \\ 
        Number of Coils & 2-24 & 2-24 & 2-28 & 2-44 \\ 
        \midrule \midrule
        \multicolumn{5}{@{}l}{\textbf{GE Data Set}}\\
        Field of View (mm) & 220$\times$220 - 240$\times$240 & 220$\times$220 - 240$\times$240 & 220$\times$220 - 240$\times$240 & 220$\times$220 - 240$\times$240 \\ 
        Matrix Size & 300$\times$300& 300$\times$300& 256$\times$256 - 320$\times$320 & 256$\times$256 - 320$\times$320 \\ 
        Slice Thickness (mm) & 5 & 5 & 5 & 5 \\ 
        Number of Slices & 8-17 & 8-17 & 10-19 & 8-19 \\ 
        TR (ms) & 2885-3268 & 2884-3268 & 3112-8400 & 3137-8400 \\ 
        TE (ms) & 24-28 & 24-28 & 95-112 & 87-107 \\ 
        Number of Coils & 12-19 & 12-19 & 12 & 12 \\ 
        \midrule \midrule
        \multicolumn{5}{@{}l}{\textbf{Philips Data Set}}\\
        Field of View (mm) & 229$\times$183 - 224$\times$224 & & 229$\times$182 - 223$\times$223 & 229$\times$186 - 224$\times$224 \\ 
        Matrix Size & 256$\times$204 - 320$\times$320 & & 248$\times$248 - 384$\times$384 & 248$\times$248 - 320$\times$320 \\ 
        Slice Thickness (mm) & 4-5 & & 4-5 & 4-5 \\ 
        Number of Slices & 15-19 & & 15-23 & 15-27 \\ 
        TR (ms) & 276-599 & & 3526-6000 & 9000-11000 \\ 
        TE (ms) & 3-10 & & 90-110 & 90-120 \\ 
        Number of Coils & 12-14 & & 12-32 & 12-32 \\ 
        \bottomrule
    \end{tabular}
    \label{tab:dataset_summary}
\end{table*}

All T2 and FLAIR Siemens scans used turbo spin echo. The T1 and T1POST Siemens scans used either turbo spin echo or FLASH. The T2 and FLAIR Philips scans used turbo spin echo. The T1 Philips scans used either spin echo or fast field echo. All GE scans used fast spin echo, i.e., were derivatives of their sequence type "FSE-XL". The T1 and T1POST GE scans were of type FLAIR.

\section{Analysis of Background Masking}
\begin{figure}[htb]
    \centering
    \includegraphics[width=0.99\textwidth]{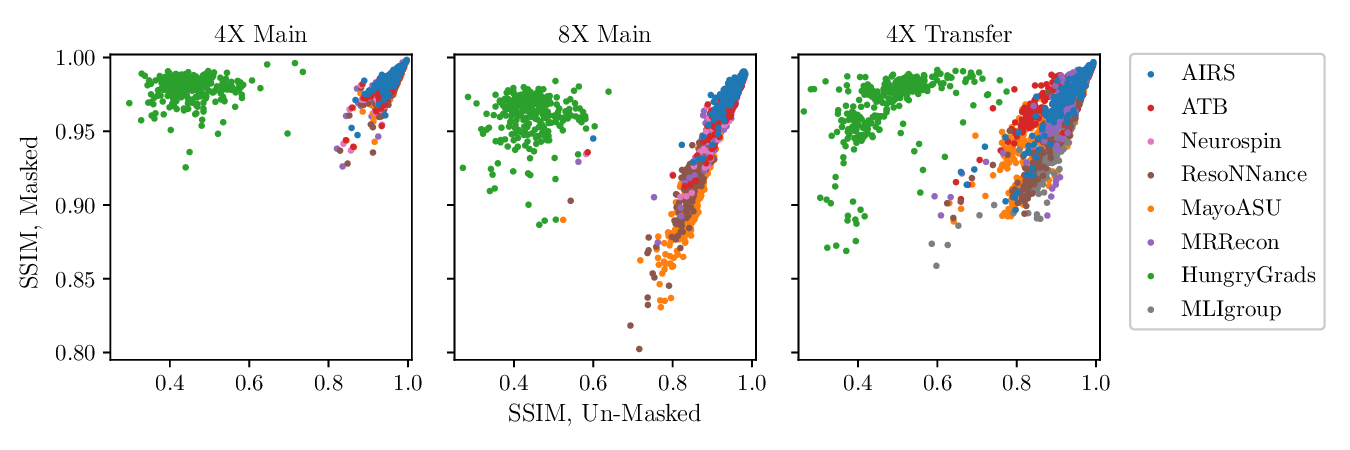}
    \caption{Overview of the impact of a masking procedure. Shown are SSIM scores incorporating masking vs. SSIM scores with no masking. Both methods used the RSS ground truth. 6 outlier points with very low SSIM on both axes were cut off for presentation in the "4X Transfer" plot.}
    \label{fig:masked_SSIM_scatter}
\end{figure}

The HungryGrads submitted to all tracks and received a very low SSIM score between 0.4 and 0.5.
The reason for the low score was setting the background to all 0s.
We performed a post-hoc analysis where we masked the challenge set prior to calculating SSIM.
The challenge set was small enough as to admit manual inspection of the masks.
The results are shown in Figure \ref{fig:masked_SSIM_scatter}.
The use of masking substantially improved the scores of HungryGrads.
However, even with the use of masking, they would not have been a finalist in any of the tracks.

\clearpage
\bibliographystyle{ieeetr}
\bibliography{references}